\documentclass[12pt]{article}
\pdfoutput=1

\usepackage[utf8]{inputenc}
\usepackage[left=2.55cm, right=2.55cm, top=2.55cm, bottom=2.55cm]{geometry}
\usepackage{amsmath,amssymb,amsbsy}
\usepackage{slashed}
\usepackage{xcolor}
\usepackage{graphicx}
\usepackage{url}
\usepackage{cancel}
\usepackage{cite}
\usepackage[colorlinks=true,allcolors=darkpurple,pdfborder={0 0 0},linktocpage=false]{hyperref}
\usepackage{tabularx,booktabs}
\usepackage{multicol}
\usepackage{feynmp}
\usepackage{units}
\usepackage{xspace}
\usepackage[labelfont=bf]{caption}
\usepackage[section]{placeins}
\usepackage{subcaption}
\usepackage{soul} 

\definecolor{darkred}{rgb}{0.6,0,0}
\definecolor{darkpurple}{rgb}{0.5,0,0.5}

\def\hc{\text{h.c.}}

\def\z2{$\mathbb{Z}_2$}
\def\321{$\mathrm{SU(3)_c} \times \mathrm{SU(2)_L} \times \mathrm{U(1)_Y}$}
\def\one{\ensuremath{\mathbf{1}}}
\def\two{\ensuremath{\mathbf{2}}}
\def\three{\ensuremath{\mathbf{3}}}
\def\threeS{\ensuremath{\mathbf{\bar 3}}}
\providecommand{\abs}[1]{\lvert#1\rvert} 

\definecolor{vdrgreen}{rgb}{0.0, 0.7, 0.0}
\definecolor{avblue}{rgb}{0.0, 0.0, 0.8}

\newcommand{\AddrIFIC}{%
  Instituto de F\'{i}sica Corpuscular, CSIC-Universitat de Val\`{e}ncia, 46980 Paterna, Spain}

\newcommand{\AddrFISTEO}{%
  Departament de F\'{\i}sica Te\`{o}rica, Universitat de Val\`{e}ncia, 46100 Burjassot, Spain}

\begin{document}

\vspace*{-2cm}
\begin{flushright}
IFIC/21-19 \\
\vspace*{2mm}
\end{flushright}

\begin{center}
\vspace*{15mm}

\vspace{1cm}
{\Large \bf 
Dark matter in a charged variant of the Scotogenic model
} \\
\vspace{1cm}

{\bf Valentina De Romeri$^{\text{a}}$, Miguel Puerta$^{\text{a}}$, Avelino Vicente$^{\text{a,b}}$}

 \vspace*{.5cm} 
 $^{(\text{a})}$ \AddrIFIC \\\vspace*{.2cm} 
 $^{(\text{b})}$ \AddrFISTEO

 \vspace*{.3cm} 
\href{mailto:deromeri@ific.uv.es}{deromeri@ific.uv.es}, \href{mailto:miguel.puerta@ific.uv.es}{miguel.puerta@ific.uv.es}, \href{mailto:avelino.vicente@ific.uv.es}{avelino.vicente@ific.uv.es}
\end{center}

\vspace*{10mm}
\begin{abstract}\noindent\normalsize
Scotogenic models are among the most popular possibilities to link dark matter and neutrino masses. In this work we discuss a variant of the Scotogenic model that includes charged fermions and a doublet with hypercharge $3/2$. Neutrino masses are induced at the one-loop level thanks to the states belonging to the dark sector. However, in contrast to the standard Scotogenic model, only the scalar dark matter candidate is viable in this version. After presenting the model and explaining some particularities about neutrino mass generation, we concentrate on its dark matter phenomenology. We show that the observed dark matter relic density can be correctly reproduced in the usual parameter space regions found for the standard Scotogenic model or the Inert Doublet model. In addition, the presence of the charged fermions may open up new regions, provided some tuning of the parameters is allowed.
\end{abstract}

\section{Introduction}
\label{sec:intro}

The Standard Model (SM) of particle physics fails to address two of
the most important open questions in current fundamental physics: the
origin of neutrino masses and the nature of dark matter (DM). While
these two issues might be completely independent, and the explanation
to the DM puzzle might even come from a completely different branch of
physics, it is tempting to explore extensions of the SM in which they
are simultaneously addressed.

Many radiative neutrino mass models are good examples of such
extensions. In this class of models, pioneered
in~\cite{Zee:1980ai,Cheng:1980qt,Zee:1985id,Babu:1988ki}, neutrino
masses vanish at tree-level but become non-zero once loop corrections
are included. This naturally explains the smallness of neutrino
masses, which get suppressed by the usual loop factors. In many cases,
they are induced at one-loop, but there are many well-known
examples leading to neutrino masses at higher loop orders,
see~\cite{Cai:2017jrq} for a review. A symmetry is often introduced to
avoid the generation of neutrino masses at tree-level. Interestingly,
the lightest state charged under this symmetry becomes completely
stable and hence can be a viable DM candidate, provided it has the
correct quantum numbers~\cite{Restrepo:2013aga}.

One of the most popular models with this feature is the so-called
Scotogenic model~\cite{Ma:2006km}. This economical setup extends the
SM particle content with three singlet fermions and one \textit{inert}
scalar doublet, all odd under a $\mathbb{Z}_2$ parity. Neutrino masses
are induced at the one-loop level and the lightest $\mathbb{Z}_2$-odd
particle, which might be a fermion or a scalar state, becomes stable
and can play the role of DM candidate. Both possibilities have been
studied in detail and shown to be valid options. The fermion DM
candidate can be produced in the early Universe via its Yukawa
interactions. Although the bounds from lepton flavor violating (LFV)
processes~\cite{Toma:2013zsa} strongly limit the allowed parameter
space in this case, the observed DM relic density can be
achieved~\cite{Vicente:2014wga} (see
also~\cite{Kubo:2006yx,Sierra:2008wj,Suematsu:2009ww,Adulpravitchai:2009gi}). Qualitatively
similar conclusions were recently found in a variant of the Scotogenic
model with Dirac fermion DM~\cite{Hagedorn:2018spx}. The scalar DM
candidate does not suffer from this limitation since it can be
produced via gauge and/or scalar interactions. In this case the DM
phenomenology resembles that of the Inert Doublet
model~\cite{Deshpande:1977rw,Barbieri:2006dq,LopezHonorez:2006gr,Honorez:2010re,Diaz:2015pyv}.

While the original Scotogenic model is a very attractive and
economical setup, it is also interesting to consider variations with a
richer phenomenology.~\footnote{The Scotogenic model can be
generalized in many different ways, see~\cite{Escribano:2020iqq} and
references therein. Furthermore, the Scotogenic mechanism can be used
to induce a small one-loop mass for a light DM
candidate~\cite{Ma:2021eko}.} For instance, a simple extension
including both singlet and triplet fermions~\cite{Hirsch:2013ola}
already leads to novel phenomenological
signatures~\cite{Rocha-Moran:2016enp,Avila:2019hhv}. In this work we
consider a variant of the Scotogenic model originally introduced
in~\cite{Aoki:2011yk}. In addition to the usual inert doublet, in this
case one introduces charged fermions and a doublet with hypercharge
$3/2$. As a consequence, the particle spectrum contains many
new charged states, including a doubly-charged scalar. The presence of
these states naturally leads to a much richer collider
phenomenology~\cite{Aoki:2011yk}. The aim of our work is to explore
the DM phenomenology of the model. In contrast to the minimal
Scotogenic model, only the scalar dark matter candidate is viable in
this version, since the rest of the $\mathbb{Z}_2$-odd states are
electrically charged. We will show that the observed DM relic
density can be correctly reproduced in this model and identify the
regions of the parameter space where this is achieved.

The rest of the manuscript is organized as follows. The model is
introduced in Sec.~\ref{sec:model}, where detailed discussions on the
scalar sector, the neutrino mass generation mechanism and the DM
candidate can be found. Sec.~\ref{sec:setup} describes the approach
followed in our numerical analysis. This Section also discusses the
most relevant experimental bounds in our setup. Our results are given
in Sec.~\ref{sec:num}, whereas a summary with the main conclusions
derived from our work is given in Sec.~\ref{sec:conclusions}. Finally,
additional details are given in Appendices~\ref{sec:app} and
\ref{sec:appyukawas}.

\section{The model}
\label{sec:model}

{
\renewcommand{\arraystretch}{1.6}
\begin{table}[tb]
\centering
\begin{tabular}{ c | c c c c c c c | c c c }
\toprule
& $q_L$ & $u_R$ & $d_R$ & $\ell_L$ & $e_R$ & $\psi_L$ & $\psi_R$ & $H$ & $\eta$ & $\Phi$ \\ 
\hline
$\rm SU(3)_C$ & $\three$ & $\threeS$ & $\threeS$ & $\one$ & $\one$ & $\one$ & $\one$ & $\one$ & $\one$ &$\one$ \\
$\rm SU(2)_L$ & $\two$ & $\one$ & $\one$ & $\two$ & $\one$ & $\one$ & $\one$ & $\two$ & $\two$ & $\two$ \\
$\rm U(1)_Y$ & $\frac{1}{6}$ & $\frac{2}{3}$ & $-\frac{1}{3}$ & $-\frac{1}{2}$ & $-1$ & $-1$ & $-1$ & $\frac{1}{2}$ & $\frac{1}{2}$ & $\frac{3}{2}$ \\[1mm]
\hline
$\mathbf{Z}_2$ & $+$ & $+$ & $+$ & $+$ & $+$ & $-$ & $-$ & $+$ & $-$ & $-$ \\
\textsc{Generations} & 3 & 3 & 3 & 3 & 3 & 2 & 2 & 1 & 1 & 1 \\
\bottomrule
\end{tabular}
\caption{Particle content of the model. $q_L$, $\ell_L$, $u_R$, $d_R$,
  $e_R$ and $H$ are the usual SM fields.
\label{tab:content}}
\end{table}
}

We consider the variant of the original Scotogenic model introduced
in~\cite{Aoki:2011yk}. The SM particle content is extended by adding
two $\rm SU(2)_L$ doublets, $\eta$ and $\Phi$, with hypercharge $1/2$
and $3/2$, respectively, and two pairs of singlet vector-like fermions
$\psi_{L,R}^a$ ($a=1,2$) with hypercharge $-1$. The scalar doublets of
the model can be decomposed into $\rm SU(2)_L$ components as
\begin{equation}
  H = \begin{pmatrix} H^+ \\ H^0 \end{pmatrix} \, , \quad
  \eta = \begin{pmatrix} \eta^+ \\ \eta^0 \end{pmatrix} \, , \quad
  \Phi = \begin{pmatrix} \Phi^{++} \\ \Phi^{+} \end{pmatrix} \, .
\end{equation}
Here $H$ is the usual SM Higgs doublet. As in the Scotogenic model, we
impose an exact $\mathbb{Z}_2$ parity. All the new particles are odd
under this symmetry, while the SM particles are assumed to be
even. The particle content of the model is summarized in
Table~\ref{tab:content}. We note that $H$ and $\eta$ are only
distinguished by their $\mathbb{Z}_2$ charges. In fact, the scalar
doublet $\eta$ has the same quantum numbers as the usual inert doublet
present in the Inert Scalar Doublet model~\cite{Deshpande:1977rw} and
the Scotogenic model~\cite{Ma:2006km}.

The most general Yukawa Lagrangian involving the new particles can be
written as
\begin{equation} \label{modlag}
    \mathcal{L}_Y = M_{\psi} \, \overline{\psi}_L \, \psi_R + Y^L \, \overline{\ell_L^c} \, \Phi \, \psi_L + Y^R \, \overline{\ell_L} \, \eta \, \psi_R + \hc \, ,
\end{equation}
where $M_{\psi}$ is a $2 \times 2$ vector-like (Dirac) mass matrix,
which we take to be diagonal without loss of generality, while $Y^L$
and $Y^R$ are dimensionless $3 \times 2$ complex matrices. The most
general scalar potential is given by
\begin{align}
  \mathcal{V} = \, & \mu^2_1 \, \abs{H}^2 + \mu^2_2 \, \abs{\eta}^2 + \mu_\Phi^2 \, \abs{\Phi}^2 + \frac{1}{2} \, \lambda_1 \, \abs{H}^4 + \frac{1}{2} \, \lambda_2 \, \abs{\eta}^4 +\frac{1}{2} \, \lambda_\Phi \, \abs{\Phi}^4 \nonumber  \\
  +& \lambda_3 \, \abs{H}^2 \, \abs{\eta}^2 + \lambda_4 \, \abs{H^\dagger\eta}^2 + \rho_1 \, \abs{H}^2 \, \abs{\Phi}^2 + \rho_2 \, \abs{\eta}^2 \, \abs{\Phi}^2 + \sigma_1 \, \abs{H^\dagger\Phi}^2 + \sigma_2 \, \abs{\eta^\dagger\Phi}^2 \label{eq:pot} \\
  +& \frac{1}{2} \left[\lambda_5 \, (H^\dagger \eta)^2 + \hc \right] + \left[\kappa \, (\Phi^\dagger H)(\eta H)+ \hc \right] \, , \nonumber
\end{align}
where $\mu_1$, $\mu_2$ and $\mu_\Phi$ are parameters with dimension
of mass and $\lambda_j$ ($j=1,2,3,4,5$), $\lambda_\Phi$, $\rho_1$,
$\rho_2$, $\sigma_1$, $\sigma_2$ and $\kappa$ are dimensionless.  We
note that in the presence of a non-zero $\kappa$ term it is not
possible to define a conserved lepton number. As shown below, this
plays a crucial role for the generation of neutrino masses.

\subsection{Scalar sector}
\label{subsec:scalar}

Let us now discuss the resulting scalar particle content of the
model. We will assume a minimum of the potential characterized by the
vacuum expectation values (VEVs)
\begin{equation}
  \langle H \rangle = \frac{1}{\sqrt{2}} \begin{pmatrix} 0 \\ v \end{pmatrix} \, , \quad \langle \eta \rangle = 0 \, , \quad \langle \Phi \rangle = 0 \, ,
\end{equation}
with $v \simeq 246$ GeV the usual SM VEV. This vacuum preserves the
$\mathbb{Z}_2$ parity, which remains unbroken after electroweak
symmetry breaking. This forbids the mixing between $H$, even under
$\mathbb{Z}_2$, and the $\eta$ and $\Phi$ doublets, odd under
$\mathbb{Z}_2$. The neutral $H^0$ and $\eta^0$ states can be split into their
CP-even and CP-odd components as
\begin{align}
  H^0 &= \frac{1}{\sqrt{2}} \left( h + i \, A + v \right) \, , \\
  \eta^0 &= \frac{1}{\sqrt{2}} \left( \eta_R + i \, \eta_I \right) \, .
\end{align}
The CP-even state $h$ can be identified with the SM Higgs boson, with
$m_h \approx 125$ GeV, while $A$ is the Goldstone boson that becomes
the longitudinal component of the $Z$ boson. Assuming that CP is
conserved in the scalar sector, $\eta_R$ and $\eta_I$ do not
mix. Their masses are given by
\begin{equation} \label{eq:mEtaRI}
  m_{\eta_{R,I}}^2 = \mu_2^2 + \frac{1}{2} \left( \lambda_3 + \lambda_4 \pm \lambda_5 \right) v^2 \, .
\end{equation}
The charged component of $H$ becomes the longitudinal component of the
$W$ boson. The charged component of $\eta$ mixes with the
singly-charged component of $\Phi$. Their Lagrangian mass term can be
written as
\begin{equation} \label{eq:massterm1}
  \mathcal{L}_m = \begin{pmatrix} \eta^- & \Phi^- \end{pmatrix} \, \begin{pmatrix} m^2_{\eta^\pm} & m^2_{12} \\ m^2_{12} & m^2_{\Phi^\pm} \end{pmatrix} \, \begin{pmatrix} \eta^+ \\ \Phi^+ \end{pmatrix} + \hc \, ,
\end{equation}
with
\begin{align}
  m^2_{\eta^\pm} &= \mu_2^2 + \frac{1}{2} \, \lambda_3 \, v^2 \, , \\
  m^2_{\Phi^\pm} &= \mu_\Phi^2 + \frac{1}{2} \, \left(\rho_1+\sigma_1\right) \, v^2 \, , \\
  m^2_{12} &= \frac{1}{2} \, \kappa \, v^2 \, .
\end{align}
The gauge eigenstates $\eta^+$ and $\Phi^+$ are related to the mass
eigenstates $H_1^+$ and $H_2^+$ as
\begin{equation} \label{eq:masseigen}
    \begin{pmatrix}
    H_1^+ \\ H_2^+
    \end{pmatrix}
    =
    \begin{pmatrix} c_\chi & s_\chi \\ -s_\chi & c_\chi
    \end{pmatrix} \,
    \begin{pmatrix} \eta^+ \\ \Phi^+
    \end{pmatrix} \, ,
\end{equation}
where $c_\chi = \cos \chi$, $s_\chi = \sin \chi$ and $\chi$ is a
mixing angle. The Lagrangian mass term can be written in terms of the
mass eigenstates $H_1^+$ and $H_2^+$ as
\begin{equation} \label{eq:massterm2}
  \mathcal{L}_m = \begin{pmatrix} H_1^- & H_2^- \end{pmatrix} \, \begin{pmatrix}  m_{H_1^\pm} & 0 \\ 0 & m_{H_2^\pm} \end{pmatrix} \, \begin{pmatrix} H_1^+ \\ H_2^+ \end{pmatrix} + \hc \, .
\end{equation}
This allows us to obtain the mixing angle $\chi$. Combining
Eqs.~\eqref{eq:massterm1} and \eqref{eq:massterm2} with
Eq.~\eqref{eq:masseigen} one finds
\begin{equation} \label{eq:chi}
    c_\chi \, s_\chi = \frac{m^2_{H^\pm_1}-m^2_{H^\pm_2}}{\left[\frac{(m^2_{\Phi^\pm}-m^2_{\eta^\pm})^2}{m^2_{12}}+4 \, m^2_{12}\right]} \approx \frac{m^2_{12}}{m^2_{H_1^\pm}-m^2_{H_2^\pm}} = \frac{\kappa \, v^2}{2 \, (m^2_{H_1^\pm}-m^2_{H_2^\pm})} \, ,
\end{equation}
where in the second step we have assumed $\chi$ to be a small mixing
angle or, equivalently, $\kappa \ll 1$. Finally, the mass of the
doubly-charged $\Phi^{++}$ is given by
\begin{equation}
  m_{\Phi^{++}}^2 = \mu_\Phi^2 + \frac{1}{2} \, \rho_1 \, v^2 \, .
\end{equation}

\subsection{Neutrino mass generation}
\label{subsec:numass}

\begin{figure}[t]
  \centering
  \includegraphics[scale=0.5]{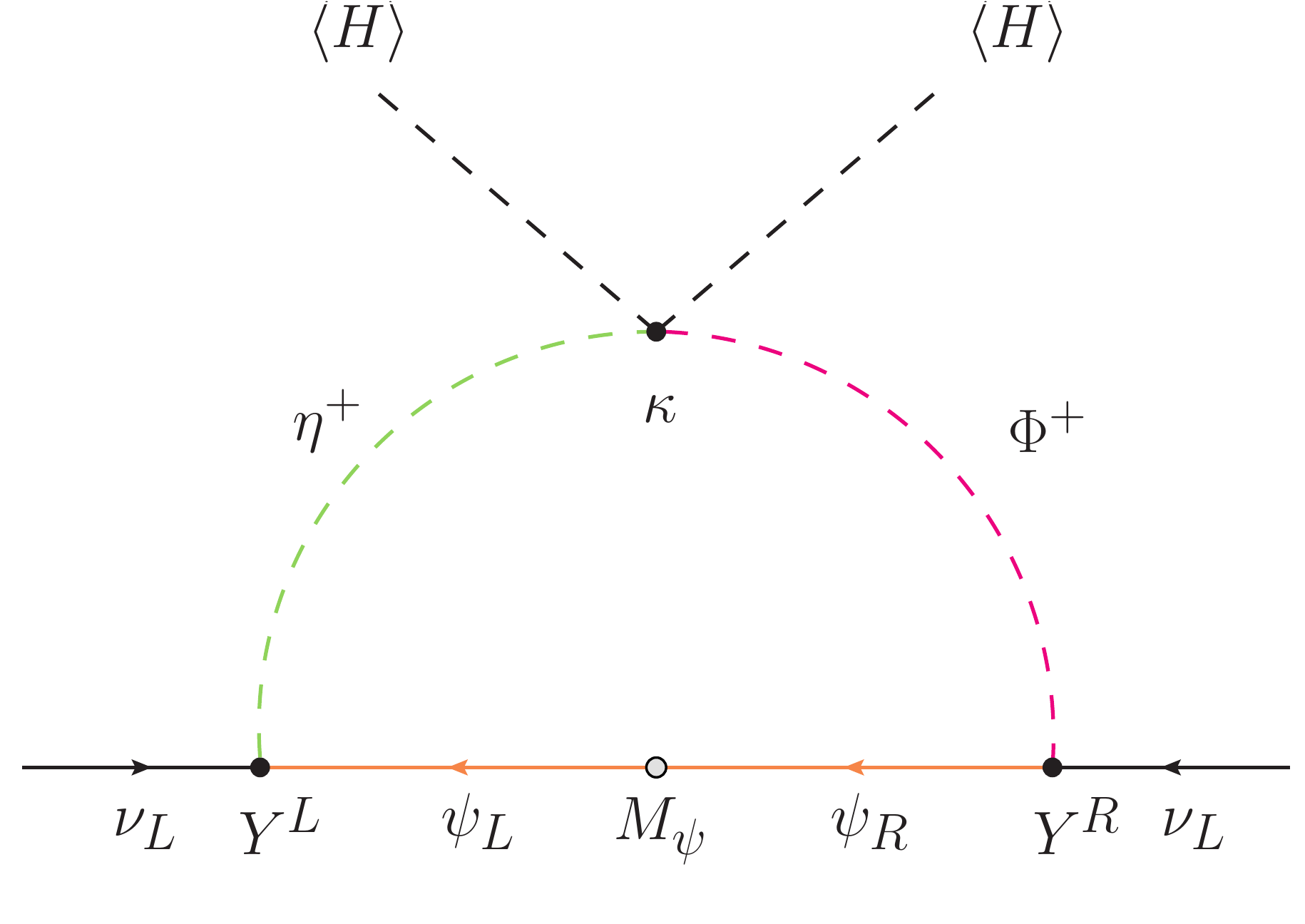}
  \caption{One-loop diagram which generates neutrino masses in the variant of the Scotogenic model under consideration. Due to electric charge conservation, in each vertex the relevant components of $\eta$ and $\Phi$ are the singly-charged ones.
  \label{fig:neumod}
  }
\end{figure}

In this model, neutrino masses are induced at the one-loop level by
the diagram in Fig.~\ref{fig:neumod}.~\footnote{All Feynman diagrams
  in our paper were made with {\tt
    JaxoDraw}~\cite{Binosi:2003yf,Binosi:2008ig}.} The
$\mathbb{Z}_2$-odd states in the loop are electrically charged, in
contrast to the original Scotogenic model, which has electrically
neutral states running in the loop. The resulting neutrino mass matrix
can be easily computed with the general expressions in
Appendix~\ref{sec:app}. In this model, $n_S = n_F = 2$, with $S \equiv
\{ H_1^\pm , H_2^\pm \}$ and $F \equiv \{\psi_1 , \psi_2 \}$. The
masses of the scalars in the loop are $m^2_{H^\pm_1}$ and
$m^2_{H^\pm_2}$, defined in Eq.~\eqref{eq:massterm2}, whereas the
masses of the fermions are $m_{\psi^1}$ and $m_{\psi^2}$, with
$m_{\psi^b}$ ($b=1,2$) the diagonal components of the matrix $M_\psi$,
introduced in Eq.~\eqref{modlag}. One can also read the couplings of
$H^\pm_1$ and $H^\pm_2$ from Eqs.~\eqref{modlag} and
\eqref{eq:masseigen}, obtaining
\begin{align}
Y_{H^\pm_1}^L &\equiv Y^L \, s_\chi \, , \quad Y_{H^\pm_1}^R \equiv Y^L \, c_\chi \, , \\
Y_{H^\pm_2}^L &\equiv Y^R \, c_\chi \, , \quad Y_{H^\pm_2}^R \equiv -Y^R \, s_\chi \, .
\end{align}
Therefore, the contribution to $(m_\nu)_{\alpha \beta}$ of the diagram
displayed in Fig.~\ref{fig:neumod} is proportional to $Y^L_{\alpha b}
\, Y^R_{\beta b}$, with $b$ the fermion index. It is easy to realize
that there is a \textit{mirror} contribution proportional to
$Y^L_{\beta b} \, Y^R_{\alpha b}$, obtained after exchanging the
scalars. In summary, the neutrino mass matrix can be written as
\begin{equation}
(m_\nu)_{\alpha \beta} = \sum_{b=1}^2 \left[ (m_\nu)^{\{H_1^\pm,\psi^b\}}_{\alpha \beta} + (m_\nu)^{\{H_2^\pm,\psi^b\}}_{\alpha \beta} \right] \, ,
\end{equation}
with
\begin{align}
    (m_\nu)^{\{H_1^\pm,\psi^b\}}_{\alpha \beta} = \frac{Y^L_{\alpha b} \, Y^R_{\beta b} + Y^R_{\alpha b} \, Y^L_{\beta b}}{16\pi^2} \, s_\chi \, c_\chi \, m_{\psi^b} \, \left[\Delta_\epsilon+1-\frac{m^2_{\psi^b} \, \log m^2_{\psi^b}-m^2_{H_1^\pm}\log m^2_{H_1^\pm}}{m^2_{\psi^b}-m^2_{H_1^\pm}}\right] \, , \label{eq:h1} \\
    (m_\nu)^{\{H_2^\pm,\psi^b\}}_{\alpha \beta} = - \frac{Y^L_{\alpha b} \, Y^R_{\beta b} + Y^R_{\alpha b} \, Y^L_{\beta b}}{16\pi^2} \, s_\chi \, c_\chi \, m_{\psi^b} \, \left[\Delta_\epsilon+1-\frac{m^2_{\psi^b}\log m^2_{\psi^b}-m^2_{H_2^\pm}\log m^2_{H_2^\pm}}{m^2_{\psi^b}-m^2_{H_2^\pm}}\right] \, . \label{eq:h2}
\end{align}
The addition of both contributions cancels out the divergence
($\Delta_\epsilon$), as expected. We can finally use
Eq.~\eqref{eq:chi} to replace the mixing angle $\chi$ in
Eqs.~\eqref{eq:h1} and \eqref{eq:h2}. By doing this, and also by
assuming the hierarchy $m^2_{H_1^\pm},m^2_{H_2^\pm} \ll m^2_{\psi^b}$,
one finds
\begin{align} \label{eq:mnufinal}
    (m_\nu)_{\alpha \beta} = \sum_{b=1}^2 \frac{Y^L_{\alpha b} \, Y^R_{\beta b} + Y^R_{\alpha b} \, Y^L_{\beta b}}{32 \, \pi^2 \, m_{\psi^a}} \frac{\kappa v^2}{m^2_{H^\pm_2}-m^2_{H^\pm_1}}\left(m^2_{H_2^\pm} \log\frac{m^2_{\psi^b}}{m^2_{H_2^\pm}}-m^{2}_{H_1^\pm} \log\frac{m^2_{\psi^b}}{m^2_{H_1^\pm}}\right) \, .
\end{align}
This expression matches the result in~\cite{Aoki:2011yk}. Neutrino
masses are proportional to the $\kappa$ parameter, as expected from
symmetry arguments. In the limit $\kappa \to 0$ lepton number is
restored, and this explains that Majorana neutrino masses can only be
induced with $\kappa \neq 0$. Furthermore, $\kappa \ll 1$ is natural,
in the sense of 't Hooft~\cite{tHooft:1979rat}. As in the Scotogenic
model, small neutrino masses can be naturally generated in this
model. One can generate $m_\nu\sim 1$ eV with $Y^{R,L} \sim 1$,
$m_{H_1^\pm}\sim 300$ GeV, $m_{H^\pm_2}\sim 400$ GeV and
$m_{\psi^a}\sim 1$ TeV if $\kappa\sim 10^{-12}$.

\subsection{Dark matter candidate}
\label{subsec:DM}

As a consequence of the conservation of the $\mathbb{Z}_2$ parity, the
lightest $\mathbb{Z}_2$-odd state is stable and cannot decay. In
contrast to the Scotogenic model, the only potentially viable
candidate is one of the neutral $\eta$ scalars, either $\eta_R$ or
$\eta_I$, since the other $\mathbb{Z}_2$-odd states are electrically
charged. Their mass difference is controlled by the $\lambda_5$
quartic coupling, as shown in Eq.~\eqref{eq:mEtaRI},
\begin{equation}
  m_{\eta_R}^2-m_{\eta_I}^2 = \lambda_5 \, v^2 \, .
\end{equation}
Another difference with respect to the Scotogenic model is that this
mass difference can in principle be large. This is because the
$\lambda_5$ coupling does not enter the neutrino mass formula, see
Eq.~\eqref{eq:mnufinal}, and can be $\sim \mathcal{O}(1)$. Also, we
note that the sign of $\lambda_5$ determines the DM
candidate. $\eta_I$ is the DM candidate if $\lambda_5 > 0$, while
$\lambda_5 < 0$ selects $\eta_R$ as the DM candidate.

The $\eta_R$ and $\eta_I$ fields couple to the electroweak gauge
bosons, since they originate from an $\rm SU(2)_L$ doublet with
hypercharge $1/2$. Their production in the early Universe is thus
expected to be generally dominated by gauge interactions. The
exception to this rule will be found when the mass of the DM candidate
is close to $\sim m_h/2$, when the so-called Higgs portal will be the
most important channel. In general, the DM phenomenology of the model
is expected to be similar to that of the Inert Scalar Doublet
model~\cite{Deshpande:1977rw}, a popular and economical model for
DM~\cite{Barbieri:2006dq,LopezHonorez:2006gr}. Finally, we point out
that the DM candidates, $\eta_R$ and $\eta_I$, carry the quantum
numbers of a supersymmetric sneutrino, except for lepton number.

\section{Analysis and experimental bounds}
\label{sec:setup}

We now proceed to discuss the approach followed in our numerical
analysis and the experimental bounds considered. The first step has
been the implementation of the model in \texttt{SARAH (version
  4.11.0)}~\cite{Staub:2013tta}, a {\tt Mathematica} package for the
analytical evaluation of the
model.~\footnote{See~\cite{Vicente:2015zba} for a pedagogical
introduction to the use of {\tt SARAH}.} With the help of this tool,
we have created a \texttt{SPheno (version
  4.0.2)}~\cite{Porod:2003um,Porod:2011nf} module with the required
numerical routines to obtain the particle spectrum and compute several
observables of interest in our model. This includes the calculation of
flavor violation observables with {\tt
  FlavorKit}~\cite{Porod:2014xia}. Finally, we have used
\texttt{micrOmegas (version 5.0.9)}~\cite{Belanger:2018ccd} for the
evaluation of DM observables, such as the DM relic density and direct
and indirect detection predictions.  We performed a numerical scan
with $\sim 10000$ points. Our choice of parameters is summarized in
Tab.~\ref{tab:inputparameters}. In particular, we choose the negative
sign for $\lambda_5$ so that throughout our analysis $\eta_R$ plays
the role of the dark matter.  Let us recall that this is a choice just
for definiteness, equivalent results would be obtained by assuming
$\eta_I$ as the lightest $\mathbb{Z}_2$-odd state. Furthermore, we
take $\mu_2^2 = \mu^2_\Phi$ in most of the parameter space to
guarantee that $\eta_R$ (and not one of the charged states from
$\Phi$) is the lightest $\mathbb{Z}_2$-odd particle. An analogous
reason lies behind the range chosen for $\mu^2_\Phi$. \\

 \begin{table}[t]
    \centering
    \begin{tabular}{|c||c|}
    \hline
         $\lambda_1 = 0.26 $ & $m_{\psi^1} = 2.1$ TeV \\
         $\lambda_2 = 0.5$ & $m_{\psi^2 }= 2.3$ TeV \\
         $\lambda_3 = 10^{-2}$ & $\rho_1 = 0.5$ \\
         $\lambda_4 \in \left[-0.5,-10^{-4} \right]$ & $\rho_2=0.7$ \\
         $\lambda_5 \in \left[-0.32,-0.003 \right]$ & $\sigma_1 \in \left[10^{-5},0.16 \right]$ \\
         $\lambda_\Phi = 3 \times 10^{-3}$ & $\sigma_2 =  10^{-2}$ \\
         $\mu^2_\Phi \in \left[100, 4.4\times 10^6 \right]$ GeV$^2$ & $\kappa= 10^{-8}$ \\
         $\mu_2^2 = \mu^2_\Phi$ except if $m_{\eta_R} \in [50, 100]$ GeV& \\
        \hline
    \end{tabular}
    \caption{Values of the input parameters considered in the numerical scan.
    \label{tab:inputparameters}}
\end{table}

The new particles introduced in this variant of the Scotogenic model may lead to different experimental signatures and affect the SM prediction of several observables. For this reason and throughout our analysis we have considered a list of experimental constraints.

\paragraph{Neutrino oscillation data} 
The generation and smallness of neutrino masses is one of the main motivations behind the idea of Scotogenic models. In our analysis, we demand compatibility of the neutrino oscillation parameters with the most recent neutrino oscillation global fit~\cite{deSalas:2020pgw}. This is achieved by adjusting the entries of the Yukawa matrices $Y^L$ and $Y^R$ by means of the master parametrization~\cite{Cordero-Carrion:2018xre,Cordero-Carrion:2019qtu}, which allows one to write
\begin{align}
  \left( Y^L \right)^T &= \frac{1}{\sqrt{2 \, f}} \, \Sigma^{-1/2} \, W \, A \, \bar{D}_{\sqrt{m}} \, U^\dagger \, , \label{eq:y1} \\
  \left( Y^R \right)^T &= \frac{1}{\sqrt{2 \, f}} \, \Sigma^{-1/2} \, W^\ast \, B \, \bar{D}_{\sqrt{m}} \, U^\dagger\, , \label{eq:y2}
\end{align}
where $\Sigma$ is a diagonal $2 \times 2$ matrix containing the
positive singular values of the matrix $M$, with $M_{ij} = \omega_i \,
\delta_{ij}$ and
\begin{equation}
  \omega_i = \frac{1} {m_{\psi^i}} \left( m^2_{H_2^\pm} \, \log\frac{m^2_{\psi^i}}{m^2_{H_2^\pm}}-m^{2}_{H_1^\pm} \, \log\frac{m^2_{\psi^i}}{m^2_{H_1^\pm}} \right) \, .
\end{equation}
Moreover,
\begin{equation}
  f = \frac{\kappa \, v^2}{32 \, \pi^2 \, (m^2_{H^\pm_2}- m^2_{H^\pm_1})}
\end{equation}
is a global factor, $U$ is the leptonic mixing matrix, a unitary $3
\times 3$ matrix that brings the neutrino mass matrix to a diagonal
form as
\begin{equation}
  U^T \, m_\nu \, U = \text{diag}(m_1,m_2,m_3) \, ,
\end{equation}
while the matrix $\bar{D}_{\sqrt{m}}$ is defined as
\begin{equation} \label{eq:defDsqrtm}
\bar{D}_{\sqrt{m}} = \left\{ \begin{array}{ll}
\text{diag}\left(\sqrt{m_1},\sqrt{m_2},\sqrt{m_3}\right) \quad & \text{if} \ m_1 \neq 0 \, , \\
P \cdot \text{diag}\left(\sqrt{v},\sqrt{m_2},\sqrt{m_3}\right) \cdot P \quad & \text{if} \ m_1 = 0 \, .
\end{array} \right.
\end{equation}
Here $v$ can actually be replaced by any arbitrary scale and $P$
depends on the neutrino mass hierarchy. In case of normal ordering,
$P$ is just the identity matrix, while for inverted ordering $P$ is a
permutation matrix that exchanges the first and third
elements. Finally, the matrices $W$, $A$ and $B$ are defined in
Appendix~\ref{sec:appyukawas}, where explicit analytical results for
the elements of the $Y^L$ and $Y^R$ matrices are also given. In our
numerical analysis we assume a normal mass ordering for light active
neutrino masses and take vanishing CP violating phases for
simplicity. We also assume that the lightest neutrino is massless.

\paragraph{Lepton flavor violation} 
While this model could in principle produce signals at facilities looking for LFV processes, these have not been observed yet. Hence we can use current null searches for LFV to constrain the parameters of the model, in particular $\kappa$ which determines the magnitude of the Yukawa matrices $Y^L$ and $Y^R$. We consider the following most stringent bounds on rare LFV processes: BR$(\mu \to e \gamma) < 4.2 \times 10^{-13}$~\cite{TheMEG:2016wtm},  BR$(\mu \to e e e)  < 1. \times 10^{-12}$~\cite{Bellgardt:1987du} and CR$(\mu^-, {\rm Au} \to e^-, \rm Au) < 7 \times 10^{-13}$~\cite{Bertl:2006up}.

\paragraph{Electroweak precision observables} 
The experimental accuracy of electroweak observables can be sensitive to the presence of new particles, like those introduced in this model. The main contribution to the higher-order calculation of electroweak precision observables is parameterized via the $\delta \rho$ parameter. We require an adequately small deviation of the $\rho$ parameter from one through the following prescription: $-0.00022 \lesssim \delta \rho \lesssim 0.00098$~\cite{Zyla:2020zbs}  ($3\sigma$ range).

\paragraph{Dark matter searches} 
We assume our DM candidate $\eta_R$ to be in thermal equilibrium with the SM particles in the early Universe. We also assume a standard cosmological scenario. If no other dark matter candidates are present, then the relic abundance of $\eta_R$ must be in agreement with the latest observations by the Planck satellite~\cite{Aghanim:2018eyx}, which set a limit on the cosmological content of cold dark matter: $ 0.1164 \leq \Omega_{\eta_R} h^2 \leq  0.1236$ (3$\sigma$ range). The relic abundance of $\eta_R$ can also be subdominant, i.e. $\Omega_{\eta_R} h^2 <  0.1164$, however in this case another DM candidate is required to explain the totality of the cosmological dark matter. Moreover,  $\eta_R$ can be probed at dark matter experiments like direct detection facilities. We apply the most stringent bound  on the WIMP-nucleon spin-independent (SI) elastic scatter cross-section from the XENON1T experiment~\cite{Aprile:2018dbl}. We compute the direct detection cross section at tree-level. Since a more exhaustive study of this observable is out of the scope of this paper, we have considered the constraint
\begin{equation}
  \frac{1}{2}\abs{\lambda_3+\lambda_4+\lambda_5} \geq 10^{-3} \, ,
\end{equation}
obtained for the inert Higgs doublet~\cite{Klasen:2013btp} in order to avoid sizable loop corrections. If $\eta_R$ annihilates into SM particles with a cross section typical of WIMPs, it may also be detected indirectly. We consider both $\gamma$ rays and antiprotons as annihilation products and compare with the respective current bounds on the WIMP annihilation cross section set by the Fermi Large Area Telescope (LAT) satellite~\cite{Ackermann:2015zua}, the ground-based arrays of Cherenkov telescopes H.E.S.S.~\cite{Abdallah:2016ygi} and the Alpha Magnetic Spectrometer (AMS-02)~\cite{Aguilar:2016kjl,Reinert:2017aga,Cuoco:2017iax} onboard the International Space Station.

\paragraph{LHC searches} The new charged particles in the model can be copiously produced and detected at the LHC, see~\cite{Aoki:2011yk} for a discussion. It is however beyond the scope of this work to perform a detailed collider study of the model. In order to guarantee compatibility with the current bounds, we have chosen $m_{\psi^{1,2}} > 1$ TeV in our numerical scan. The singly-charged scalars $H_{1,2}^+$ have masses in a wide range of values, always in the hundreds of GeV. Finally, the doubly-charged scalar $\Phi^{++}$ is chosen to be always heavier than $\sim 200$ GeV by imposing $\mu_2^2 \neq \mu^2_\Phi$ in the region $m_{\eta_R}\in\left[50\text{ GeV},100\text{ GeV}\right]$. This may seem as too little restrictive, since the currently most stringent bounds on a doubly-charged scalar range from $220$ GeV~\cite{Aaboud:2018qcu} (for a $\Phi^{++}$ that decays into $W^+ W^+$) to $846$ GeV~\cite{Aaboud:2017qph} (for a $\Phi^{++}$ that decays into $\mu^+ \mu^+$). However, notice that in our model $\Phi^{++}$ is $\mathbb{Z}_2$-odd and its decays always include DM particles. In fact, we have found that in many parameter points $\Phi^{++}$ decays as $\Phi^{++} \rightarrow H_1^+ \, W^+$, followed by $H_1^+ \to \eta_R \, \ell^+ \, \nu$, thus leading to large amounts of missing energy in the final state.

\paragraph{Invisible decay width of the Higgs boson}
If the new neutral scalars $\eta_R$ and $\eta_I$ are light enough, new
invisible decay channels of the Higgs boson will be kinematically
accessible. We require that their contribution to the invisible decay
width of the Higgs boson is not larger than the currently most
stringent experimental bound: BR$(h \to \rm inv) \lesssim
19\%$~\cite{Sirunyan:2018owy}. Our computation of BR$(h \to
\rm inv)$ includes the decay BR$(h \to \eta_R \, \eta_R)$, when
kinematically possible, and also BR$(h \to \eta_I \, \eta_I)$ when
$m_{\eta_I}-m_{\eta_R} < 10$ GeV. We estimate that for larger mass
differences between $\eta_I$ and $\eta_R$, $\eta_I$ would decay
visibly inside the detector. Moreover, the presence of charged scalars
may also modify the Higgs photon coupling to two photons. For this
reason we further impose that $0.84\lesssim$ BR$(h \to \gamma
\gamma)/$BR$(h \to \gamma \gamma)_{\rm SM} \lesssim
1.41$, with BR$(h \rightarrow \gamma \gamma)_{\rm SM} = (0.227 \pm 0.011)\times 10^{-2} $~\cite{Zyla:2020zbs}.

\paragraph{$\boldsymbol{\eta_I}$ lifetime}
Searches for long-lived particles at the LHC usually look for
displaced vertices. In order to ensure compatibility with the
non-observation of such signatures, we require $\tau (\eta_I) \ll 1$
s. This translates into the constraint $\abs{\lambda_5} \geq 5 \times
10^{-4}$, which guarantees that $\eta_I$ decays fast enough into
$\eta_R$.

\paragraph{Theoretical considerations} 
We require that the expansion of the scalar potential (see Eq.~\ref{eq:pot}) around its minimum must be perturbatively valid. At this scope, we take all scalar quartic couplings to be $\lesssim 1$. Similarly, the elements of the $Y^L$ and $Y^R$ Yukawa matrices are also bounded to be $\lesssim 1$. \\

We summarize in table~\ref{tab:constraints} the list of experimental constraints applied in the numerical scan.

 \begin{table}[!htb]
    \centering
    \begin{tabular}{|c|c|}
    \hline
         BR$(\mu \rightarrow e \gamma)$ & $< 4.2 \times 10^{-13} $~\cite{TheMEG:2016wtm}  \\  
BR$(\mu\rightarrow e e e)  $ & $< 1. \times 10^{-12} $~\cite{Bellgardt:1987du}  \\ 
CR$(\mu^-, {\rm Au} \to e^-, \rm Au)$ & $ < 7 \times 10^{-13}$~\cite{Bertl:2006up}\\
$\frac{\text{BR}(h  \rightarrow \gamma \gamma )}{\text{BR}(h \rightarrow \gamma \gamma )_{\rm SM}} $ &  $[0.84, 1.41]$~\cite{Zyla:2020zbs}  \\
BR$(h \rightarrow \gamma \gamma)_{\rm SM}$ & $ (0.227 \pm 0.011)\times 10^{-2} $~\cite{Zyla:2020zbs}  \\
BR$(h \rightarrow \rm inv)$ & $ 0.19 $~\cite{Sirunyan:2018owy}  \\
$\delta \rho$ & $ [-0.00022, 0.00098]$~\cite{Zyla:2020zbs} \\
$\Omega_{\eta_R} h^2$  & $ [0.1164, 0.1236]$~\cite{Zyla:2020zbs} \\
        \hline
    \end{tabular}
    \caption{List of experimental constraints applied to the numerical scan. }
    \label{tab:constraints}
\end{table}

\section{Numerical results}
\label{sec:num}

In this section we summarize our results of the analysis of $\eta_R$ as a dark matter candidate in the model. As we previously commented, this is the only potential DM candidate in this model, given the choice of negative sign of $\lambda_5$ (see Sec.~\ref{subsec:DM}). \\

\begin{figure}[!hbt]
  \centering
  \includegraphics[width=0.8\linewidth]{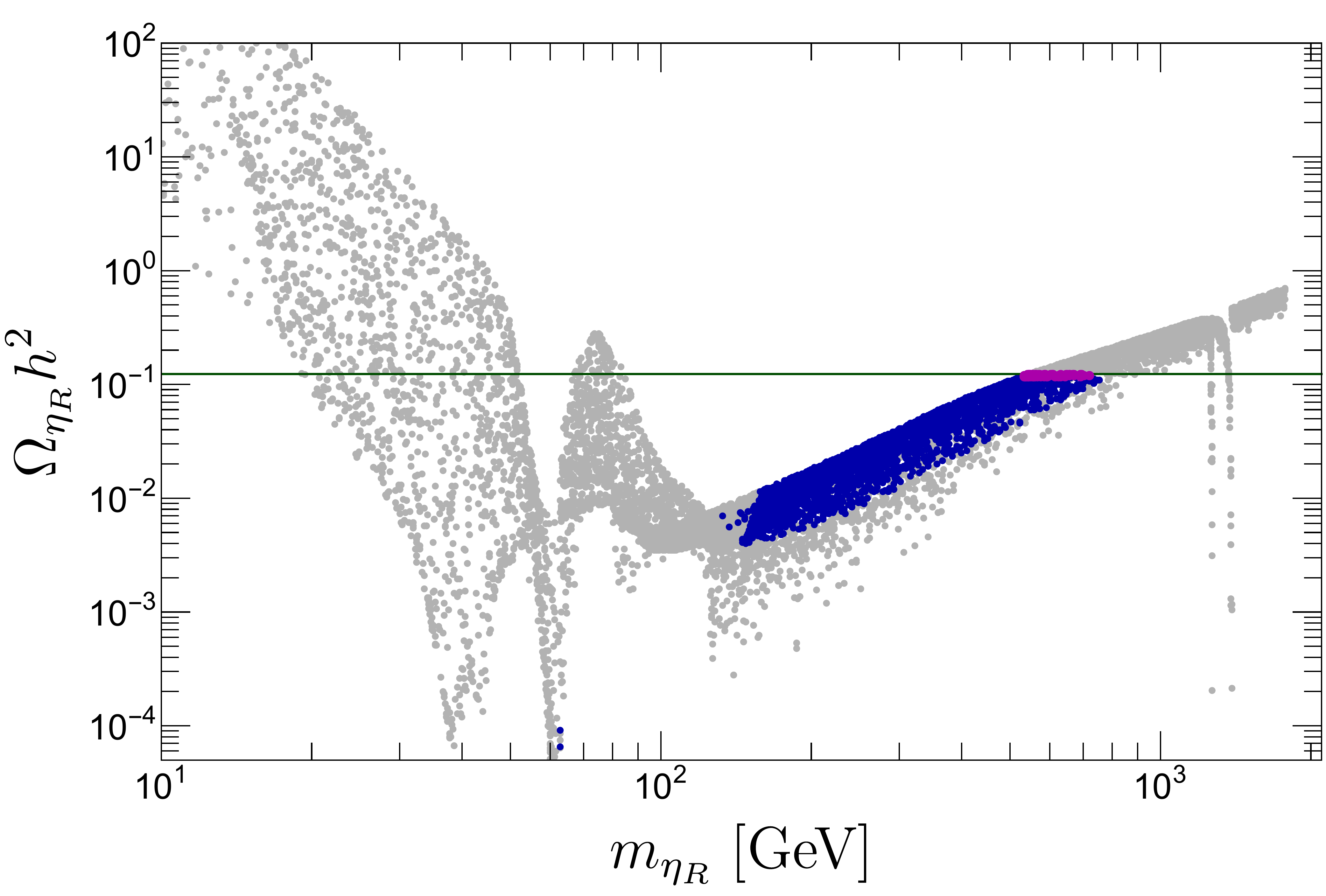}
  \caption{Relic abundance of $\eta_R$ as a function of $m_{\eta_R}$.  Magenta points depict solutions in agreement with the cold dark matter measurement obtained from Planck data~\cite{Aghanim:2018eyx} (green band, 3$\sigma$ interval). Blue points denote allowed solutions but leading to underabundant dark matter. Gray points are excluded by any of the constraints listed in Sec.~\ref{sec:setup}.
  \label{fig:omega-metaR}}
\end{figure}

Fig.~\ref{fig:omega-metaR} shows the expected $\eta_R$ relic abundance as a function of the mass of $\eta_R$. Points colored in magenta denote solutions which can reproduce the observed cold dark matter relic density, as they fall within the 3$\sigma$ range derived by the Planck satellite data~\cite{Aghanim:2018eyx}, $\Omega_{\eta_R} h^2 = 0.120 \pm 0.0036$ (green band). From this analysis we can identify that the preferred DM mass range lies around $500~{\rm GeV} \lesssim m_{\eta_R} \lesssim 800$ GeV.
Blue points also refer to allowed solutions but where $\eta_R$ would be a subdominant component of dark matter, and another candidate would be required. Finally, gray points denote solutions excluded by any of the constraints previously discussed in Sec.~\ref{sec:setup}. In particular, the Planck constraint itself rules out most of the solutions leading to excessively large relic density. Another large set of solutions with $70~{\rm GeV} \lesssim m_{\eta_R} \lesssim 700$ GeV is excluded by the current bound on WIMP-nucleon SI elastic scattering cross section set by the experiment XENON1T~\cite{Aprile:2018dbl}.

Looking at the figure from left to right, the first dip refers to the $Z$-pole, where $m_{\eta_R} \sim M_Z/2$ and annihilations via s-channel $Z$ exchange are relevant. A similar feature appears at $m_{\eta_R} \sim 60$ GeV, where a narrower dip indicates efficient annihilations via s-channel Higgs exchange. However, most solutions with $m_{\eta_R} \lesssim 60$ GeV appear to be in conflict with current collider limits on BR$(h \to \rm inv)$ and with LHC searches for doubly-charged scalars and are therefore excluded. Notice that the s-channel Higgs annihilations are in general more efficient than the $Z$-mediated ones, not being momentum suppressed, and in principle could lead to solutions with $\Omega_{\eta_R} h^2 < 10^{-5}$. Only very few solutions eventually survive in the Higgs-pole region at $m_{\eta_R} \lesssim 100$ GeV, due to a variety of constraints, among which BR$(h \to \gamma \gamma)$ and BR$(h \to \rm inv)$ are the most important ones. These parameter points have very small relic density.
After the Higgs pole, $\eta_R$ quartic interactions with gauge bosons and heavy quarks become effective, when kinematically allowed.  In particular, $\eta_R$ annihilations into $W^+ W^-$ via quartic couplings are efficient at $m_{\eta_R} \gtrsim 80$ GeV, translating into a third drop in the relic abundance. As soon as kinematically allowed, $\eta_R$ can then annihilate also into two Higgs bosons. In general, for  $m_{\eta_R} \gtrsim 100$ GeV, the $\eta_R$ relic density increases as its annihilation cross section drops proportionally as $\sim \frac{1}{m_{\eta_R}^2}$.
 
\begin{figure}[!hbt]
  \centering
  \includegraphics[width=0.6\linewidth]{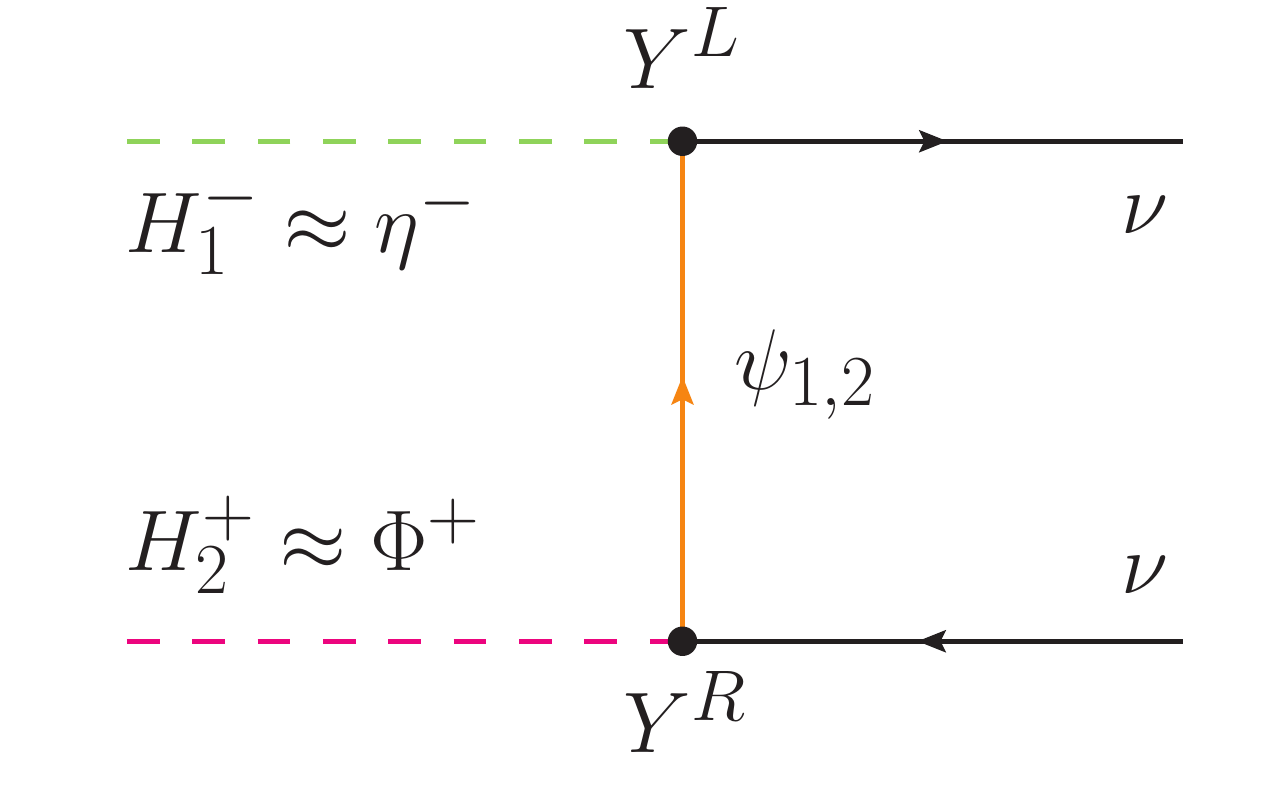}
  \caption{Dominant diagram contributing to the co-annihilation channel $H_1^- H_2^+ \rightarrow \nu \nu$. Notice that both Yukawa matrices, $Y^L$ and $Y^R$, must have sizable entries to make this process efficient.
  \label{fig:DM_t_channel}}
\end{figure}

\begin{figure}[!h]
  \centering
  \includegraphics[width=0.48\linewidth]{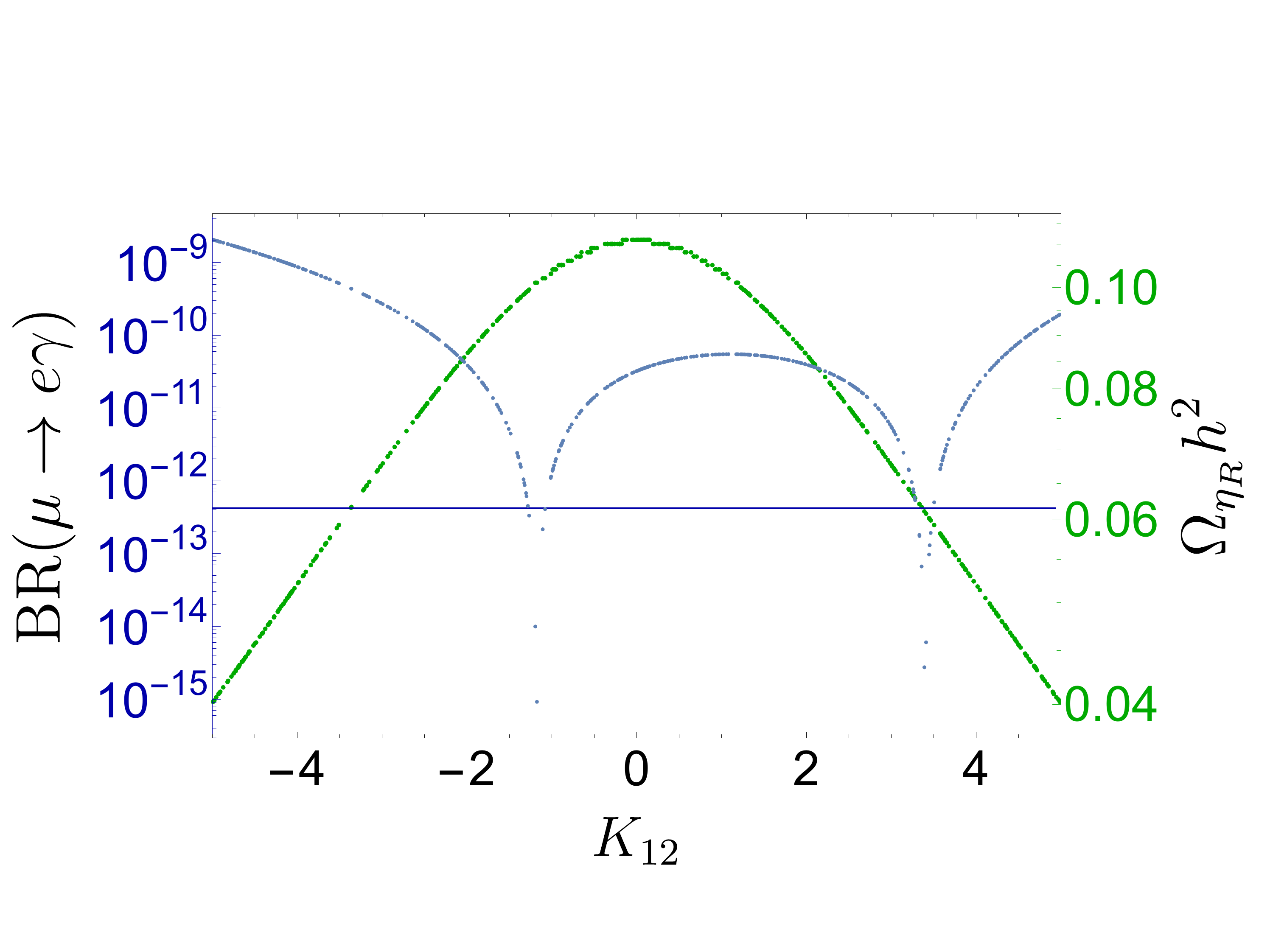}
  \includegraphics[width=0.48\linewidth]{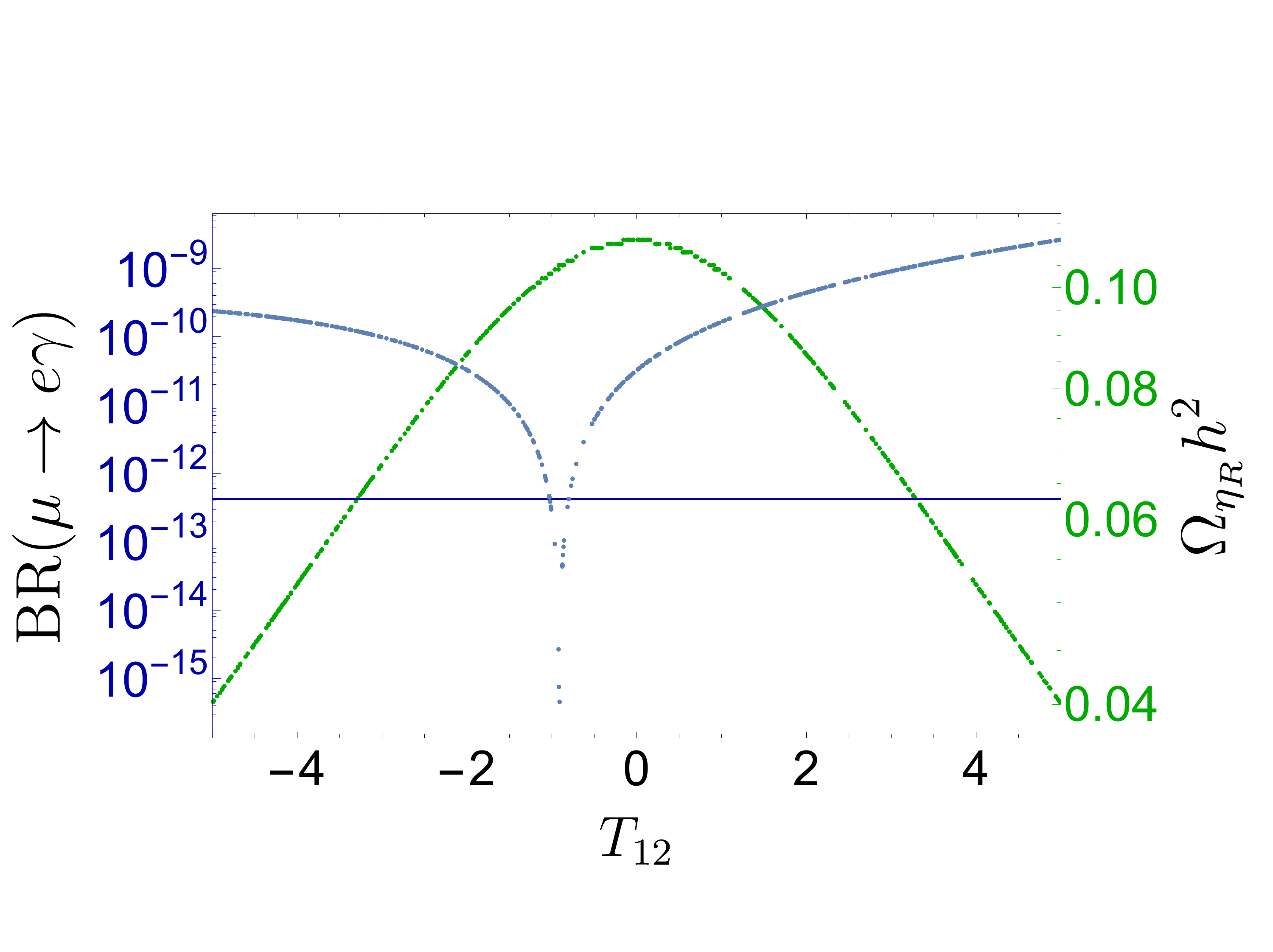}
  \caption{BR($\mu \rightarrow e \gamma$) (left vertical axis, blue points) and relic abundance of $\eta_R$ (right vertical axis, green points) as a function of $K_{12}$ (left panel) and $T_{12}$ (right panel). The horizontal line represents the current upper bound on BR($\mu \rightarrow e \gamma$)~\cite{TheMEG:2016wtm}. Therefore, all blue points above this line are excluded.
  \label{fig:LFV_K12_T12}}
\end{figure}

A very interesting feature appears at $m_{\eta_R} \sim 1.3$ TeV. The relic density suddenly drops to very small values due to a very efficient co-annihilation channel $H_1^- H_2^+ \rightarrow \nu \nu$, mediated by the vector-like fermions $\psi_{1,2}$ in t-channel, as shown in Fig.~\ref{fig:DM_t_channel}. This contribution is made very efficient by the magnitude of the relevant entries of the Yukawa matrices entering the diagram. The same Yukawas also play a relevant role in the rare LFV process $\mu \rightarrow e \gamma$, which turns out to exceed the current limit set by MEG~\cite{TheMEG:2016wtm} and thus excludes all solutions falling in this very narrow pole. Nonetheless, given the large freedom due to the vast array of parameters that can be varied in the parameterization of the Yukawa matrices (see Appendix~\ref{sec:appyukawas}), one can always find a fine-tuned combination which allows to keep the co-annihilation channel $H_1^- H_2^+ \rightarrow \nu \nu$ efficient while at the same time not leading to a BR($\mu \rightarrow e \gamma$) in conflict with current observations. This can be seen in  Fig~\ref{fig:LFV_K12_T12}, where we show the dependence of BR($\mu \rightarrow e \gamma$) and  $\Omega_{\eta_R} h^2$ as a function of the parameters $K_{12}$ (left panel) and $T_{12}$ (right panel). To obtain these figures we extracted one solution at $m_{\eta_R} \sim 1.3$ TeV from our general scan (see Fig.~\ref{fig:omega-metaR}) --- which we recall was made fixing $K_{12}=0$ and $T_{12}=0$ --- and we scanned around those central values.  Eventually, few allowed solutions (both with a relic abundance matching current observations and under-abundant) will appear at  $m_{\eta_R} \sim 1.3$ TeV. Let us notice that similar features with fine-tuned solutions may be present also in other parts of the parameter space.\\

\begin{figure}[!hbt]
  \centering
  \includegraphics[width=0.8\linewidth]{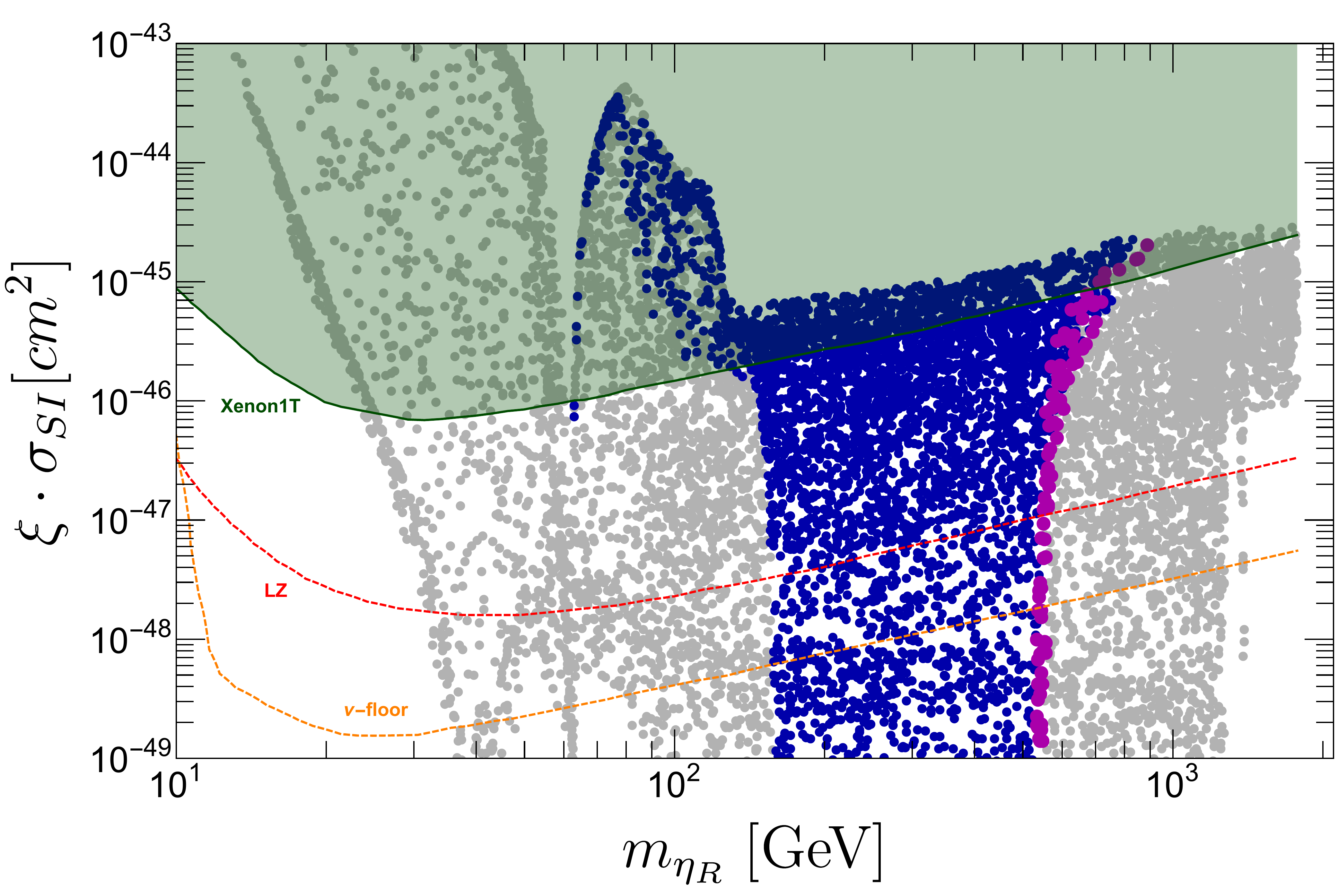}
  \caption{Spin-independent $\eta_R$-nucleon elastic scattering cross section -- weighted by the relative abundance -- versus $m_{\eta_R}$. Same color code as in Fig.~\ref{fig:omega-metaR}. The green dashed area is excluded by the most recent upper bound from the XENON1T experiment~\cite{Aprile:2018dbl}. The dashed orange curve indicates the expected discovery limit corresponding to the ``$\nu$-floor" from CE$\nu$NS of solar and atmospheric neutrinos for a Ge target~\cite{Billard:2013qya}.  The dashed red curve depicts the projected sensitivity expected at LUX-ZEPLIN (LZ)~\cite{Akerib:2018lyp}. 
  \label{fig:DD}}
\end{figure}

Next we proceed to discuss the results for $\eta_R$ direct detection.\\
In full generality, the tree-level SI $\eta_R$-nucleon interaction cross section receives two contributions, from scattering through the Higgs or $Z$ bosons. However, since the $\lambda_5$ quartic coupling induces a mass splitting between $\eta_R$ and $\eta_I$, eventually it turns out that the interaction through the $Z$ boson is kinematically forbidden, or leads to inelastic scattering, in a large part of the parameter space. Should this not be the case, the $\eta_R$ - nucleon interaction through the $Z$ boson would dominate (since the $\eta$ doublet has hypercharge different from zero) and very likely exceed the current bounds from DM direct detection experiments.
We show in Fig.~\ref{fig:DD} the SI $\eta_R$-nucleon elastic scattering cross section, weighted by $\xi =  \frac{\Omega_{\eta_R}}{\Omega_{\rm CDM,Planck}}$ versus the $\eta_R$ mass.  We use the same color code as in Fig.~\ref{fig:omega-metaR}. The plain green line and dashed area indicate the current most stringent limit from XENON1T~\cite{Aprile:2018dbl}.  Even if not shown here, other stringent constraints  on the SI $\eta_R$-nucleon elastic scattering cross section apply from the liquid xenon experiments LUX~\cite{Akerib:2016vxi,Akerib:2018hck} and PandaX-II~\cite{Cui:2017nnn}. Liquid argon experiments like DarkSide-50~\cite{Agnes:2018fwg} and DEAP-3600~\cite{Ajaj:2019imk} are instead presently limited and hence less constraining, due to lower exposures and the currently low acceptance in DEAP-3600.
We also illustrate for comparison the expected discovery limit corresponding to the ``$\nu$-floor" from coherent elastic neutrino-nucleus scattering (CE$\nu$NS)  for a Ge target~\cite{Billard:2013qya} (dashed orange line), as well as the sensitivity projection (90\% CL) for the future experiment LUX-ZEPLIN (LZ, red dashed)~\cite{Akerib:2018lyp}. Other future experiments like XENONnT~\cite{Aprile:2020vtw}, DarkSide-20k~\cite{DS_ESPP}, ARGO~\cite{DS_ESPP} and DARWIN~\cite{Schumann:2015cpa,Aalbers:2016jon}  (see~\cite{Billard:2021uyg} for a recent review), not shown here not to overcrowd the figure, will in principle be able to explore the parameter space of this model. 
Given the current experimental constraints, most of the solutions with a relic abundance falling within the $3\sigma$ C.L. cold dark matter range obtained by the Planck collaboration~\cite{Aghanim:2018eyx} lie in a narrow region around $m_{\eta_R} \sim 500-700$ GeV. Many more solutions leading to under-abundant dark matter lie at $\sim 150~\text{GeV} \lesssim m_{\eta_R} \sim 500$ GeV and could be tested at future DD experiments. On the whole, the phenomenology of the scalar DM candidate in this model, while richer, shares similar properties to that of the analogous scalar DM candidates in the simple Scotogenic Model~\cite{Ma:2006km}, in Inert Scalar Doublet models, and in other Scotogenic variants like the Singlet-Triplet Scotogenic model~\cite{Avila:2019hhv}.\\

Another $\eta_R$ search strategy is via its indirect detection.  If it annihilates into SM particles or messengers (with sizable annihilation cross section, close to the thermal value) it can contribute to the flux of cosmic particles that reaches the Earth. Photons, and more specifically $\gamma$ rays are among the most suitable messengers to probe WIMP dark matter indirectly. They are not deflected during propagation, so they carry information about their source, and they are relatively easy to detect. When $\eta_R$ is lighter than $M_W$, only annihilations into fermions lighter than $m_{\eta_R}$ are allowed at tree level, with the heaviest kinematically allowed fermion final states dominating (i.e. $\eta_R \eta_R \rightarrow b \bar{b}$ and $\eta_R \eta_R \rightarrow \tau^+ \tau^-$). At higher masses, when kinematically allowed through the corresponding quartic couplings, also the following channels open: $\eta_R \eta_R \rightarrow W^+ W^-, h h, Z^0 Z^0$.
The hadronization of the gauge bosons, Higgs boson or quarks produces neutral pions, which in turn decay into photons thus giving rise to a $\gamma$-ray flux with a continuum spectrum which may be detected at indirect detection experiments.
Besides this featureless $\gamma$-ray spectrum, the model also predicts a spectral feature from the internal bremsstrahlung process $\eta_R \eta_R \rightarrow W^+ W^- \gamma$, similarly to the Inert Doublet model~\cite{Garcia-Cely:2013zga,Giacchino:2013bta}. We include this contribution in our analysis.

While more challenging due to uncertainties in the treatment of their propagation, charged particles can also be used to probe for $\eta_R$ annihilations. PAMELA~\cite{Adriani:2008zr,Adriani:2013uda} and, more recently, the AMS-02~\cite{Accardo:2014lma,Aguilar:2019owu} positrons data allow to place constraints on annihilating WIMPs, which are particularly stringent if they annihilate mainly to the first two generations of charged leptons.  In our case, at $m_{\eta_R} < m_W$, $\eta_R$ annihilates predominantly into $\tau$s (and $b\bar{b}$). Hence, bounds from cosmic positrons are less stringent than those from $\gamma$ rays. On the other hand, AMS-02 data on the antiproton flux and the Boron to Carbon (B/C) ratio can be used to constrain the $\eta_R$ annihilation cross section~\cite{Aguilar:2016kjl,Reinert:2017aga,Cuoco:2017iax}. Provided that  astrophysical uncertainties on the $\bar{p}$ production, propagation and on Solar modulation are reliably taken into account (see e.g.~\cite{Cuoco:2019kuu,Cholis:2019ejx,Heisig:2020nse}), these bounds result to be stronger than $\gamma$-ray limits from dwarf spheroidal satellite galaxies (dSphs) over a wide mass range.\\

\begin{figure}[!h]
  \centering
  \includegraphics[width=0.85\linewidth]{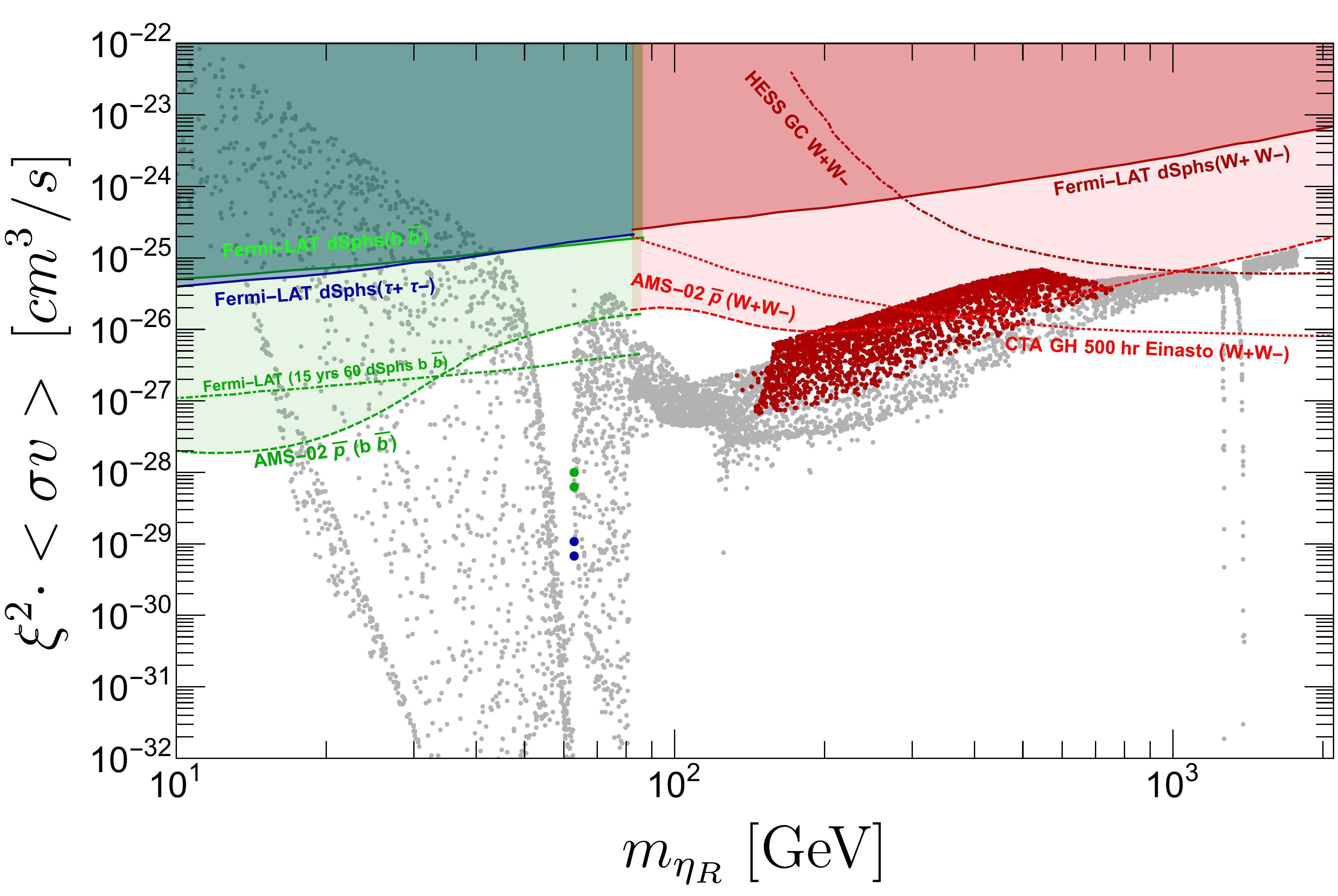}
  \caption{$\eta_R$ annihilation cross section for annihilations into $b \bar{b}$ (green), $\tau^+ \tau^-$ (blue) and $W^+ W^-$ (red) final states. The green, blue and red plain lines refer to the corresponding 95\% C.L. upper limits currently set by Fermi-LAT $\gamma$-ray data from dSphs~\cite{Ackermann:2015zua}. The dark red dot-dashed curve is the current 95\% C.L. upper limit obtained by H.E.S.S. using GC data~\cite{Abdallah:2016ygi}. The green and red dashed lines denote current 95\% C.L. constraints derived from the antiproton and B/C data of AMS-02~\cite{Reinert:2017aga}. For comparison, we also show sensitivity projections for Fermi-LAT ($b \bar{b}$ channel, assuming 60 dSphs and 15 years of data)~\cite{Charles:2016pgz} and for CTA (looking at the galactic halo, $W^+ W^-$ channel)~\cite{CTAConsortium:2018tzg}.
\label{fig:ID}}
\end{figure}

We focus on the main $\eta_R$ annihilation channels into $b \bar{b}$, $\tau^+ \tau^-$ and $W^+ W^-$ to compare with current limits set by $\gamma$-ray observations of Milky Way dSphs with Fermi-LAT data~\cite{Ackermann:2015zua}, of the Galactic Center (GC) with the H.E.S.S. array of ground-based Cherenkov telescopes~\cite{Abdallah:2016ygi} and a combination of $\bar{p}$  and B/C data of AMS-02~\cite{Aguilar:2016kjl,Reinert:2017aga}.
Fig.~\ref{fig:ID} shows the results of our numerical scan of the $\eta_R$ annihilation cross section (weighted by $\xi^2$ and by the correspondent branching ratio) versus its mass, for the main annihilation channels:  $\eta_R \eta_R \rightarrow b \bar{b}$ (green points), $\eta_R \eta_R \rightarrow \tau^+ \tau^-$ (blue points) and $\eta_R \eta_R \rightarrow W^+ W^-$ (dark red points). As in previous plots, gray points denote solutions excluded by any of the constraints listed in Sec.~\ref{sec:setup}. In the same figure we depict the 95\% C.L. upper limits currently set by the Fermi-LAT with $\gamma$-ray observations of Milky Way dSphs (6 years, Pass 8 event-level analysis)~\cite{Ackermann:2015zua} (plain curves and shaded areas, assuming annihilation into $b \bar{b}$ (green), $\tau^+ \tau^-$ (blue) and $W^+ W^-$ (dark red)). 
We also show the current upper limit obtained with data accumulated over 10 years by H.E.S.S. observations of the GC~\cite{Abdallah:2016ygi}, assuming a $W^+ W^-$ annihilation channel and an Einasto dark matter density profile (dark red dot-dashed curve and shaded area).  Current bounds from a combination of $\bar{p}$  and B/C data of AMS-02~\cite{Aguilar:2016kjl,Reinert:2017aga} are instead shown as dashed curves (green for $b \bar{b}$ and red for $W^+ W^-$ channels, respectively). 
As for comparison we further illustrate projected sensitivities for Fermi-LAT from a stacked analysis of 60 dSphs and 15 years of data, in the $b \bar{b}$ channel~\cite{Charles:2016pgz}  (dot-dashed green line) and for the Cherenkov Telescope Array (CTA), for the Milky Way galactic halo target, assuming the $W^+ W^-$ annihilation channel and an Einasto dark matter density profile~\cite{CTAConsortium:2018tzg}.  
Our model predictions, not excluded by other constraints, lie at least a factor of few below current limits set by $\gamma$-ray observations with Fermi-LAT and H.E.S.S..
In particular, allowed solutions in the ``low" (and narrow) mass window around $m_{\eta_R} \sim 60$ GeV ($b \bar{b}$ and $\tau^+ \tau^-$ annihilation channels) are hardly found, so only a couple of them are shown for illustration. Predictions for these channels lie well below current bounds.
 Antiprotons and B/C data of AMS-02 instead already allow to exclude some solutions with $200~\text{GeV} \lesssim m_{\eta_R} \lesssim 800$ GeV ($W^+ W^-$ annihilation channel). We remark that these bounds are obtained under significant astrophysical uncertainties. Future telescopes like CTA as well as additional LAT data will allow to further explore these regions of the parameter space, thus allowing for multi-messenger indirect probes of this model.
Finally let us comment that the annihilation cross section of non-relativistic $\eta_R$ at the current epoch can be affected by a non-perturbative correction, the Sommerfeld enhancement~\cite{1931AnP...403..257S,Hisano:2003ec,Hisano:2004ds,ArkaniHamed:2008qn,Chowdhury:2016mtl} which would consequently affect also its indirect detection signatures. However, we did not include it in our calculation as a thorough treatment of this effect lies beyond the scope of this paper.

\section{Conclusions}
\label{sec:conclusions}

We have considered a variant of the Scotogenic model with an extended scalar content, in which one of the new doublets carries hypercharge $Y= 3/2$. As a consequence, the model particle spectrum contains new states, including a doubly-charged scalar, singly-charged scalars and new charged fermions leading to a rich collider phenomenology. Neutrino masses are generated radiatively as in other Scotogenic scenarios. However, contrarily to the simple Scotogenic model,  in this setup only the lightest $\mathbb{Z}_2$-odd neutral scalar $\eta$ is a viable dark matter candidate.
We have analyzed the phenomenology of the dark matter particle $\eta_R$ and shown that the observed relic density can be obtained for $500~{\rm GeV} \lesssim m_{\eta_R} \lesssim 800$ GeV. This result is equivalent to that obtained in the Inert Doublet model. In addition, we have found that the correct relic density can also be achieved at larger $\eta_R$ masses if the coannihilation channel $H_1^- H_2^+ \rightarrow \nu \nu$ becomes efficient. However, this requires some fine-tuning to avoid the stringent constraints from lepton flavor violation. Moreover, we have presented a full numerical analysis of the signatures expected at dark matter detection experiments, both via direct and indirect probes. 
We have found that most recent direct detection data from XENON1T already rule out a region of the parameter space in the mass range $m_{\eta_R} \sim 100$ GeV. Finally we have commented about indirect detection signatures with $\gamma$-rays and antiprotons data. While bounds derived from $\gamma$-ray data from Fermi-LAT and H.E.S.S. currently lie above the model predictions, AMS-02 antiproton data instead already allows to constrain some solutions in the $200~{\rm GeV} \lesssim m_{\eta_R} \lesssim 800$ GeV mass range.

\section*{Acknowledgements}

Work supported by the Spanish grants FPA2017-85216-P
(MINECO/AEI/FEDER, UE), SEJI/2018/033, SEJI/2020/016 (Generalitat Valenciana) and
FPA2017-90566-REDC (Red Consolider MultiDark). AV acknowledges
financial support from MINECO through the Ramón y Cajal contract
RYC2018-025795-I. VDR acknowledges financial support by 
the Universitat de Val\`encia through the sub-programme “ATRACCI\'O DE TALENT 2019”.

\appendix

\section{Neutrino mass matrix in Scotogenic models}
\label{sec:app}

\begin{figure}[t]
  \centering
  \includegraphics[scale=0.5]{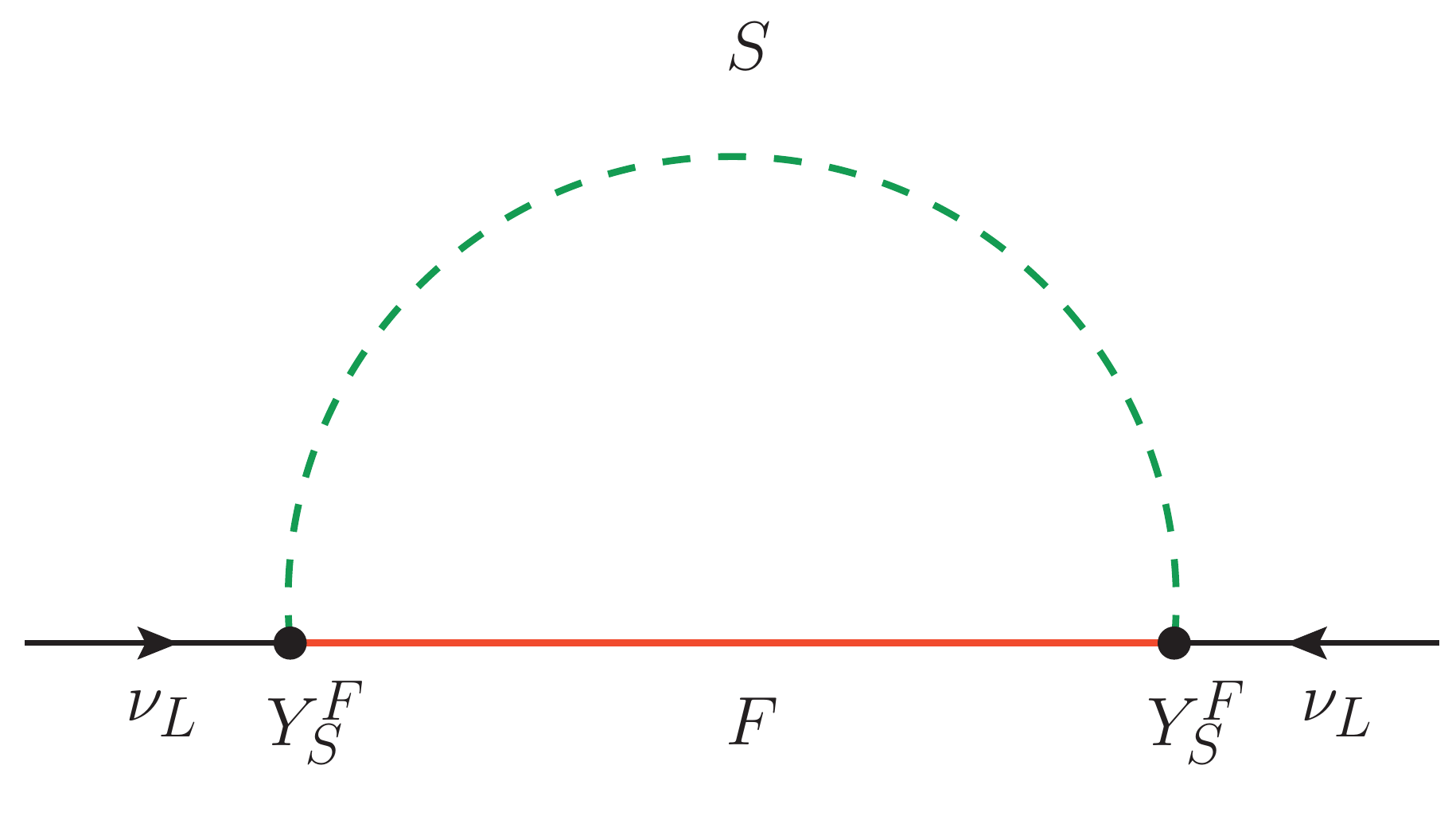}
  \caption{Generic one-loop contribution to the neutrino mass matrix. $F$ and $S$ represent fermion and scalar mass eigenstates contributing to the neutrino mass matrix.
    \label{fig:gendiag}
    }
\end{figure}

In this Appendix we find a general analytical expression for one-loop
contributions to the neutrino mass matrix in Scotogenic models. Let us
consider a generic model with $n_S$ scalars $S_a$ with masses $m_{Sa}$
($a=1,\dots,n_S$) and $n_F$ fermions $F_b$ with masses $m_{Fb}$
($b=1,\dots,n_F$) that couple to light neutrinos with interaction
terms of the form
\begin{equation} \label{genlag}
  \mathcal{L} = \left( Y_S^F \right)_{\alpha a b} \, \overline{\nu_L}_\alpha \, S_a \, F_b + \hc \, ,
\end{equation}
where $Y^F_S$ is a $3 \times n_S \times n_F$ complex object. This
interaction term induces neutrino masses at the one-loop level via the
diagram in Fig.~\ref{fig:gendiag}. The resulting neutrino mass matrix
can be expressed as
\begin{equation} \label{eq:mnu}
  i \, (m_\nu)_{\alpha \beta} = i \, \sum_{a,b} (m_\nu)^{\{S_a,F_b\}}_{\alpha \beta} \, .
\end{equation}
where the sum extends over all mass eigenstates running in the
loop. Each $i \, (m_\nu)^{\{S_a,F_b\}}_{ij}$ represents the individual
contribution by the scalar $S_a$ and the fermion $F_b$ and is given by
\begin{equation}
  i \, (m_\nu)^{\{S_a,F_b\}}_{\alpha \beta} = (Y_S^F)_{\alpha a b} \, (Y_S^F)_{\beta a b} \, \int \frac{d^Dk}{(2\pi)^D} \frac{(\slashed{k}+m_{Fb})}{(k^2-m^2_{Sa})(k^2-m^2_{Fb})} \, ,
\end{equation}
where $D = 4 - \varepsilon$ is the number of space-time dimensions,
$k$ is the momentum running in the loop and the external neutrinos
have been taken at rest. The $\slashed{k}$ piece does not contribute
since it is an odd function in $k$. The remaining term is
logarithmically divergent. Introducing the usual Feynman's
parameters, we can rewrite our expression as
\begin{equation}
  i \, (m_\nu)^{\{S_a,F_b\}}_{\alpha \beta} = (Y^F_S)_{\alpha a b} \, (Y^F_S)_{\beta a b} \, \int^1_0 dx\int\frac{d^Dk}{(2\pi)^D} \frac{m_{Fb}}{(k^2-C_{ab}^2)^2} \, ,
\end{equation}
where $C_{ab}^2=m_{Sa}^2+m^2_{Fb}(1-x)$. One can now integrate in $k$ and obtain
\begin{equation}
  i \, (m_\nu)^{\{S_a,F_b\}}_{\alpha \beta} = \frac{i \, (Y_S^F)_{\alpha a b} \, (Y_S^F)_{\beta a b}}{16\pi^2} \, m_{Fb} \, \int_0^1 dx \left[\Delta_\epsilon-\log C_{ab} +\mathcal{O}(\epsilon)\right] \, ,
\end{equation}
where $\Delta_\epsilon=\frac{2}{\epsilon}-\gamma+\log 4\pi$ and
$\gamma$ is Euler's constant. Finally, we integrate in $x$ and take
the limit $\epsilon \to 0$ (which corresponds to $D\to 4$). This leads
us to
\begin{equation} \label{eq:mnuind}
  (m_\nu)^{\{S_a,F_b\}}_{\alpha \beta} = \frac{(Y^F_S)_{\alpha a b} \, (Y_S^F)_{\beta a b}}{16\pi^2} \, m_{Fb} \, \left[\Delta_\epsilon+1-\frac{m^2_{Fb}\log(m^2_{Fb})-m^2_{Sa} \log(m^2_{Sa})}{m^2_{Fb}-m^2_{Sa}}\right] \, .
\end{equation}
Eq.~\eqref{eq:mnuind} is our \textit{master} expression for the
one-loop contributions to the neutrino mass matrix. In order to
compute the neutrino mass matrix in different versions of the Scotogenic model,
one just needs to identify the scalar and fermion mass eigenstates
running in the loop and replace the $Y^F_S$ couplings by their
expressions in terms of the model parameters. Then, summing over all
individual contributions one finds the total one-loop neutrino mass
matrix.  We note that individual contributions to $m_\nu$ are
divergent due to the $\Delta_\epsilon$ term. However, the one-loop
neutrino mass matrix is finite. Therefore, these divergences cancel
out in the sum in Eq.~\eqref{eq:mnu}.

As an example, we show this explicitly in the original Scotogenic
model~\cite{Ma:2006km}. In this case one introduces the inert scalar
doublet $\eta$, with the same definition as in the model in
Sec.~\ref{sec:model}, as well as three generations of fermions $N$,
singlets under the SM gauge group. The new $N$ and $\eta$ fields
couple to the SM lepton doublet with the Yukawa term $Y^N \,
\overline{\ell_L} \, \widetilde \eta \, N$, with $\widetilde \eta = i
\tau_2 \eta^*$ and $\tau_2$ the second Pauli matrix. Moreover, the
neutral component of the $\eta$ doublet can been split into its
CP-even and CP-odd components as
\begin{equation}
\eta^0 = \frac{1}{\sqrt{2}} \left( \eta_R + i \, \eta_I \right) \, .
\end{equation}
Assuming that CP is conserved in the scalar sector, $\eta_R$ and
$\eta_I$ do not mix and constitute mass eigenstates. Therefore, in the
Scotogenic model one has $n_S = 2$ and $n_F = 3$, with $S \equiv \{
\eta_R , \eta_I \}$ and $F \equiv \{N_1 , N_2 , N_3 \}$. Their
couplings are given by
\begin{equation}
 Y^{\eta_R} \equiv \frac{Y^N}{\sqrt{2}} \, , \quad Y^{\eta_I} \equiv i \, \frac{Y^N}{\sqrt{2}} \, .
\end{equation}
Finally, applying Eq.~\eqref{eq:mnuind} and summing over all the
states in the loop we are able to obtain a finite expression (the
divergences of the diagrams with $\eta_R$ and $\eta_I$ cancel each
other) for the neutrino mass matrix
\begin{equation}
    (m_\nu)_{\alpha \beta} = \sum_{b=1}^3 \frac{Y^N_{\alpha b} \, Y^N_{\beta b}}{32\pi^2} \, m_{Nb}\left[\frac{m^2_{\eta_R}}{m^2_{Nb}-m^2_{\eta_R}}\log\frac{m^2_{\eta_R}}{m^2_{Nb}}-\frac{m^2_{\eta_I}}{m^2_{Nb}-m^2_{\eta_I}}\log\frac{m^2_{\eta_I}}{m^2_{Nb}}\right] \, ,
\end{equation}
where $m_{\eta_R}$ and $m_{\eta_I}$ are the corresponding masses for
each component. This expression differs from the expression
in~\cite{Ma:2006km} by a factor of $1/2$, an error that was first
identified in v1 of~\cite{Merle:2015gea}.

\section{General parametrization for the Yukawa matrices}
\label{sec:appyukawas}

The master
parametrization~\cite{Cordero-Carrion:2018xre,Cordero-Carrion:2019qtu}
allows one to write the $3\times 2$ Yukawa matrices $Y^L$ and $Y^R$ in
terms of neutrino oscillation parameters as in Eqs.~\eqref{eq:y1} and
\eqref{eq:y2}. In these expressions, $W$ is a $2 \times 2$ unitary
matrix, given in terms of the complex angle $\beta$ as
\begin{equation}
  W = \begin{pmatrix}
    \cos \beta & \sin \beta \\
    - \sin \beta^* & \cos \beta^* \end{pmatrix} \, ,
\end{equation}
$A = T \, C_1$, with $T$ a $2 \times 2$ upper-triangular
matrix with positives real entries in the diagonal, and $B =
(T^T)^{-1} (C_1 C_2 + K C_1)$, with $K$ a $2 \times 2$ antisymmetric
matrix, hence containing only one degree of freedom. The matrices
$C_1$ and $C_2$ are given by
\begin{equation}
  C_1 = \begin{pmatrix}
    z_1 & 0 & 0 \\
    z_2 & 0 & 0 \end{pmatrix} \, , \quad C_2 = \begin{pmatrix}
    -1 & 0 & 0 \\
    0 & 1 & 0 \\
    0 & 0 & 1 \end{pmatrix} \, ,
\end{equation}
with $z_1$ and $z_2$ two complex numbers such that $z_1^2 + z_2^2 =
0$. With all these ingredients one can obtain explicit expressions for
the elements of $Y^L$ and $Y^R$, which will be determined from
neutrino oscillation data as well as from a set of free parameters. We
point out that these are the most general expressions for the elements
of the $Y^L$ and $Y^R$ Yukawa matrices:

\begin{multline}
  Y^L_{11} =  \frac{c_{12} c_{13} \sqrt{m_1} (c_\beta (z_1 T_{11}+z_2 T_{12})+z_2 T_{22} s_\beta )+c_{13} e^{-i \eta_2} \sqrt{m_2} s_{12} T_{11} c_\beta }{\sqrt{2f} \sqrt{\omega_1}}   \\ + \frac{\sqrt{m_3} s_{13} e^{i (\delta -\eta_3)} (T_{22} s_\beta +T_{12} c_\beta )}{\sqrt{2f} \sqrt{\omega_1}}  \, , 
\end{multline}
\begin{multline}
  Y^L_{21} =  - \frac{ \sqrt{m_1} (c_\beta (z_1 T_{11}+z_2 T_{12})+z_2 T_{22} s_\beta ) \left( c_{23} s_{12}+c_{12} e^{-i \delta } s_{13} s_{23}\right) }{\sqrt{2f} \sqrt{\omega_1}} \\ - e^{-i (\eta_2+\eta_3)} \frac{e^{i \eta_3} \sqrt{m_2} \ T_{11} c_\beta \left(c_{12} c_{23}-e^{-i \delta } s_{12} s_{13} s_{23}\right)+c_{13} e^{i \eta_2} \sqrt{m_3} s_{23} (T_{22} s_\beta +T_{12} c_\beta )}{\sqrt{2f} \sqrt{\omega_1}}  \, , 
\end{multline}
\begin{multline}
  Y^L_{31} = e^{-i (\eta_2+\eta_3)}  \frac{e^{i \eta_2} c_{13} c_{23} \sqrt{m_3} \ (T_{22} s_\beta +T_{12} c_\beta )}{\sqrt{2f} \sqrt{\omega_1}} \\
  +  \frac{\sqrt{m_1} (z_1 T_{11} c_\beta +z_2 T_{22} s_\beta +z_2 T_{12} c_\beta ) \left(s_{12} s_{23}-c_{12} c_{23} e^{-i \delta } s_{13}\right)}{\sqrt{2f} \sqrt{\omega_1}} \\ 
  + e^{-i (\eta_2+\eta_3)} \frac{- e^{i \eta_3} \sqrt{m_2} \ T_{11} c_\beta \left(c_{23} e^{-i \delta } s_{12} s_{13}+c_{12} s_{23}\right)}{\sqrt{2f} \sqrt{\omega_1}} \, ,
\end{multline}
  \begin{multline}
  Y^L_{12} = e^{-i (\eta_2+\eta_3)} \frac{ e^{i \eta_2} T_{22} c_\beta ^* \left(c_{12} c_{13} e^{i \eta_3} \sqrt{m_1} z_2+e^{i \delta } \sqrt{m_3} s_{13}\right)}{\sqrt{2f} \sqrt{\omega_2}} \\ -e^{-i (\eta_2+\eta_3)}\frac{s_\beta ^* \left(c_{13} e^{i \eta_3} \left(\sqrt{m_2} s_{12} T_{11}+c_{12} e^{i \eta_2} \sqrt{m_1} (z_1 T_{11}+z_2 T_{12})\right)+\sqrt{m_3} s_{13} T_{12} e^{i (\delta +\eta_2)}\right)}{\sqrt{2f} \sqrt{\omega_2}} \, ,
 \end{multline}
\begin{multline}
     Y^L_{22} = \frac{\sqrt{m_1} \left(c_{23} s_{12}+c_{12} e^{-i \delta } s_{13} s_{23}\right) \left(s_\beta ^* (z_1 T_{11}+z_2 T_{12})-z_2 T_{22} c_\beta ^*\right)}{\sqrt{2f} \sqrt{\omega_2}} \\ 
     -e^{-i (\eta_2+\eta_3)}\frac{e^{i \eta_3} \sqrt{m_2} \ T_{11} s_\beta ^* \left(c_{12} c_{23}-e^{-i \delta } s_{12} s_{13} s_{23}\right)-c_{13} e^{i \eta_2} \sqrt{m_3} s_{23} \left(T_{12} s_\beta ^* -T_{22} c_\beta ^*\right)}{\sqrt{2f} \sqrt{\omega_2}} \, , 
\end{multline}
\begin{multline}
     Y^L_{32} = \frac{\sqrt{m_1} \left(c_{12} c_{23} e^{-i \delta } s_{13}-s_{12} s_{23}\right) \left(s_\beta ^* (z_{1} T_{11}+z_{2} T_{12})z_{2} T_{22} c_\beta ^*\right)}{\sqrt{2f} \sqrt{\omega_2}} \\
     +\frac{e^{-i \eta 2} \sqrt{m_2} \ T_{11} s_\beta ^* \left(c_{23} e^{-i \delta } s_{12} s_{13}+c_{12} s_{23}\right)-c_{13} c_{23} e^{-i \eta 3} \sqrt{m_{3}} \left(T_{12} s_\beta ^*-T_{22} c_\beta ^*\right)}{\sqrt{2f} \sqrt{\omega_2}}  \, , 
\end{multline}
\begin{multline}
     Y^R_{11} = -e^{-i (\eta_2+\eta_3)} \frac{ s_\beta^* c_{12} c_{13} \sqrt{m_1} e^{i (\eta_2+\eta_3)} (z_1 K_{12} T_{11}+z_2 K_{12} T_{12}-z_1 T_{12}+z_2 T_{11})}{\sqrt{2f} \sqrt{\omega_1} \ T_{11} T_{22}} \\ +e^{-i (\eta_2+\eta_3)} \frac{s_\beta^*\left( -c_{13} e^{i \eta_3} \sqrt{m_2} s_{12} (K_{12} T_{11}+T_{12})+\sqrt{m_3} s_{13} e^{i (\delta +\eta_2)} (T_{11}-K_{12} T_{12})\right)}{\sqrt{2f} \sqrt{\omega_1} \ T_{11} T_{22}} \\ +e^{-i (\eta_2+\eta_3)} \frac{T_{22} c_\beta^*  \left(c_{13} e^{i \eta_3} \left(\sqrt{m_2} s_{12}-c_{12} e^{i \eta_2} \sqrt{m_1} (z_1-z_2 K_{12})\right)+\sqrt{m_3} s_{13} K_{12} e^{i (\delta +\eta_2)}\right)}{\sqrt{2f} \sqrt{\omega_1} \ T_{11} T_{22}}
  \, , 
\end{multline}
\begin{multline}
    Y^R_{21} = \frac{ s_\beta^* \sqrt{m_1} (z_1 K_{12} T_{11}+z_2 K_{12} T_{12}-z_1 T_{12}+z_2 T_{11}) \left(c_{23} s_{12}+c_{12} e^{-i \delta } s_{13} s_{23}\right)}{\sqrt{2f} \sqrt{\omega_1} \ T_{11} T_{22}} \\
    - e^{-i (\eta_2+\eta_3)} \frac{ s_\beta^*\sqrt{m_1} \left( e^{i \eta_3} \sqrt{m_2} (K_{12} T_{11}+T_{12}) \left(c_{12} c_{23}-e^{-i \delta } s_{12} s_{13} s_{23}\right)+c_{13} e^{i \eta_2} \sqrt{m_3} s_{23} (T_{11}-K_{12} T_{12})\right)}{\sqrt{2f} \sqrt{\omega_1} \ T_{11} T_{22}} \\ 
     +e^{-i (\eta_2+\eta_3)} \frac{T_{22} c_\beta^*  \left(\sqrt{m_1} e^{i (\eta_2+\eta_3)} (z_1-z_2 K_{12}) \left(c_{23} s_{12}+c_{12} e^{-i \delta } s_{13} s_{23}\right)+c_{13} e^{i \eta 2} \sqrt{m_3} s_{23} K_{12}\right)}{\sqrt{2f} \sqrt{\omega_1} \ T_{11} T_{22}} \\ + e^{-i (\eta_2+\eta_3)} \frac{T_{22} c_\beta^*  \ e^{i \eta 3} \sqrt{m_{2}} \left(c_{12} c_{23}-e^{-i \delta } s_{12} s_{13} s_{23}\right)}{\sqrt{2f} \sqrt{\omega_1} \ T_{11} T_{22}}
  \, ,
\end{multline}
\begin{multline}
Y^R_{31} = \frac{-\sqrt{m_1} \left(s_{12} s_{23}-c_{12} c_{23} e^{-i \delta } s_{13}\right) \left(s_\beta^* (T_{11} (z_1 K_{12}+z_2)-T_{12} (z_1-z_2 K_{12}))\right)}{\sqrt{2f} \sqrt{\omega_1} \ T_{11} T_{22}} \\ 
-\frac{\sqrt{m_1} \left(s_{12} s_{23}-c_{12} c_{23} e^{-i \delta } s_{13}\right)T_{22} c_\beta^*  (z_1-z_2 K_{12})}{\sqrt{2f} \sqrt{\omega_1} \ T_{11} T_{22}} \\
+\frac{e^{-i \eta_2} \sqrt{m_2} \left((K_{12} T_{11}+T_{12}) s_\beta^* -T_{22} c_\beta^* \right) \left(c_{23} e^{-i \delta } s_{12} s_{13}+c_{12} s_{23}\right)}{\sqrt{2f} \sqrt{\omega_1} \ T_{11} T_{22}} \\
+ \frac{c_{13} c_{23} e^{-i \eta_3} \sqrt{m_3} \left((T_{11}-K_{12} T_{12}) s_\beta^* +K_{12} T_{22} c_\beta ^*\right)}{\sqrt{2f} \sqrt{\omega_1} \ T_{11} T_{22}} \, ,
\end{multline}  
\begin{multline}
 Y^R_{12} = -\frac{c_\beta c_{12} c_{13} \sqrt{m_1} (z_1 K_{12} T_{11}+z_2 K_{12} T_{12}-z_1 T_{12}+z_2 T_{11})}{\sqrt{2f} \sqrt{\omega_2} \ T_{11} T_{22}} \\
 -e^{-i (\eta_2+\eta_3)} \frac{c_\beta \left( -c_{13} e^{i \eta_3} \sqrt{m_2} s_{12} (K_{12} T_{11}+T_{12})+\sqrt{m_3} s_{13} e^{i (\delta +\eta_2)} (T_{11}-K_{12} T_{12})\right)}{\sqrt{2f} \sqrt{\omega_2} \ T_{11} T_{22}} \\
 -e^{-i (\eta_2+\eta_3)} \frac{T_{22} s_\beta \left(c_{13} e^{i \eta_3} \left(\sqrt{m_2} s_{12}-c_{12} e^{i \eta_2} \sqrt{m_1} (z_1-z_2 K_{12})\right)+\sqrt{m_3} s_{13} K_{12} e^{i (\delta +\eta_2)}\right)}{\sqrt{2f} \sqrt{\omega_2} \ T_{11} T_{22}} \, ,
\end{multline}
\begin{multline}
 Y^R_{22} = \frac{\sqrt{m_1} \left(c_{23} s_{12}+c_{12} e^{-i \delta } s_{13} s_{23}\right) (T_{22} s_\beta (z_2 K_{12}-z_1)+c_\beta (z_1 K_{12} T_{11}+z_2 K_{12} T_{12}-z_1 T_{12}+z_2 T_{11}))}{\sqrt{2f} \sqrt{\omega_2} \ T_{11} T_{22}} \\
 -e^{-i (\eta_2+\eta_3)}\frac{e^{i \eta_3} \sqrt{m_2} (c_\beta (K_{12} T_{11}+T_{12})+T_{22} s_\beta ) \left(c_{12} c_{23}-e^{-i \delta } s_{12} s_{13} s_{23}\right)}{\sqrt{2f} \sqrt{\omega_2} \ T_{11} T_{22}}  \\
 +e^{-i (\eta_2+\eta_3)} \frac{c_{13} e^{i \eta_2} \sqrt{m_3} s_{23} (c_\beta (T_{11}-K_{12} T_{12})-K_{12} T_{22} s_\beta )}{\sqrt{2f} \sqrt{\omega_2} \ T_{11} T_{22}} \, ,
\end{multline}
\begin{multline}
 Y^R_{32} =\frac{-\sqrt{m_1} \left(s_{12} s_{23}-c_{12} c_{23} e^{-i \delta } s_{13}\right) (T_{22} s_\beta (z_2 K_{12}-z_1)+c_\beta (z_1 K_{12} T_{11}+z_2 K_{12} T_{12}-z_1 T_{12}+z_2 T_{11}))}{\sqrt{2f} \sqrt{\omega_2} \ T_{11} T_{22}} \\
 +e^{-i (\eta_2+\eta_3)} \frac{e^{i \eta_3} \sqrt{m_2} (c_\beta (K_{12} T_{11}+T_{12})+T_{22} s_\beta )}{\sqrt{2f} \sqrt{\omega_2} \ T_{11} T_{22}} \\
 + e^{-i (\eta_2+\eta_3)} \frac{\left(c_{23} e^{-i \delta } s_{12} s_{13}+c_{12} s_{23}\right)+c_{13} c_{23} e^{i \eta_2} \sqrt{m_3} (c_\beta (T_{11}-K_{12} T_{12})-K_{12} T_{22} s_\beta )}{\sqrt{2f} \sqrt{\omega_2} \ T_{11} T_{22}} \, .
\end{multline}
In these expressions $c_{kl} \equiv \cos\theta_{kl}$ and $s_{kl}
\equiv \sin{\theta_{kl}}$ ($k=1,2$ and $l=2,3$; $l\neq k$) are the
usual neutrino mixing angles measured in oscillation experiments,
while $m_j \equiv \sqrt{m_1^2 + \Delta m^2_{j1}}$ ($j=2,3$) are
neutrino mass eigenvalues. Moreover, $m_1$ is the lightest neutrino
mass, $\eta_i$ ($i=2,3)$ are CP violating Majorana phases and $\delta$
is the CP violating Dirac phase. We also use the notation $s_\beta
  \equiv \sin \beta$ and $c_\beta \equiv \cos \beta$. In our
numerical scans we have fixed $T_{11} = T_{22} = 1$ and $\beta =
T_{12} = K_{12} = z_1 = z_2 = m_1 = \eta_i = \delta = 0$ unless
otherwise stated. \\

\bibliographystyle{utphys}
\bibliography{refs}

\end{document}